\documentclass[review,times,12pt,authoryear]{elsarticle}
\usepackage[inner=2.5cm,outer=2.5cm,bottom=2.5cm]{geometry}
\usepackage{amssymb}
\usepackage{amsthm}
\usepackage{graphicx,psfrag,epsf,epstopdf}
\usepackage{enumerate}
\usepackage{siunitx}
\usepackage{natbib}
\usepackage{amsmath,amsfonts,amssymb,amsthm}
\usepackage{bm}    
\usepackage{threeparttable}   
\usepackage{tcolorbox}   
\usepackage{color,float}   
\usepackage{multi row}
\usepackage{caption}
\usepackage[toc,page]{appendix}
\usepackage{subcaption}
\usepackage{setspace}
\usepackage{hyperref}
\hypersetup{
	colorlinks=true,
	linkcolor=blue,
	citecolor=blue
}

\usepackage[figuresright]{rotating}

\newtheorem{theorem}{Theorem}

\newtheorem{proposition}{Proposition} 
  
\newtheorem{lemma}{Lemma}

\def \R {\mathbb{R}}

\def \E {\mathbb{E}}

\def\1{1\!{\rm l}}

\begin{document}

\begin{frontmatter}

\title{\bf \large Bivariate Distribution Regression with Application to Insurance Data\tnoteref{thank}}
\tnotetext[thank]{
  We would like to thank helpful comments from the editor and two anonymous
  referees. 
  We are grateful to Angelos Dassios, Jiti Gao, Constantinos Kardaras,
  Nadja Klein,
  Jonas Meier,  
  Ryo Okui,
  Peng Shi, Qiwei Yao,
	and other participants of the seminars at the London School of Economics and Political Science and the University of Melbourne, the 2021 IME conference, and the 2022 Asian Meeting of the Econometric Society in East and South-East Asia for their helpful comments and suggestions. 
	Oka gratefully acknowledges
	financial support from the Australian Government through the Australian Research Council's Discovery Projects (project DP190101152).
  }

\author[yyw]{Yunyun Wang}
\address[yyw]{Department of Econometrics and Business Statistics, Monash University }
\ead{yunyun.wang@monash.edu}
\author[tat]{Tatsushi Oka} 
\address[tat]{AI Lab, CyberAgent}
\ead{oka\_tatsushi@cyberagent.co.jp}
\author[dan]{Dan Zhu\corref{cor1}}
\address[dan]{Department of Econometrics and Business Statistics, Monash University }
\ead{dan.zhu@monash.edu}
\cortext[cor1]{Corresponding author}

\vspace{-0.5cm}
\begin{abstract}
	Understanding variable dependence, particularly eliciting their statistical properties given a set of covariates, provides the mathematical foundation in practical operations management such as risk analysis and decision-making given observed circumstances.
	This article presents an estimation method for modeling the conditional joint distribution of bivariate outcomes based on the distribution regression and factorization methods. This method is considered semiparametric in that it allows for flexible modeling of both the marginal and joint distributions conditional on covariates without imposing global parametric assumptions across the entire distribution. In contrast to existing parametric approaches, our method can accommodate discrete, continuous, or mixed variables, and provides a simple yet effective way to capture distributional dependence structures between bivariate outcomes and covariates.
	Various simulation results confirm that our method can perform
	similarly or better in finite samples compared to the alternative methods.
	In an application to the study of a motor third-party liability insurance portfolio, the proposed method effectively estimates risk measures such as the conditional Value-at-Risk and Expected Shortfall. This result suggests that this semiparametric approach can serve as an alternative in insurance risk management.   
\end{abstract}


\end{frontmatter}


\section{Introduction}	
 \label{sec:intro}  
 
Data with bivariate discrete and continuous outcomes are
often encountered in various areas, including
economics, insurance risk analysis, and production management.
Characterizing the dependencies among the outcomes
and their joint distributional features is a crucial yet challenging task due to their complexities, especially when conditioning on observed variables. In insurance data analysis,  companies maintain a record of the number of claims and the average claim amount from their policyholders of non-life insurance. 
The study of their joint distribution, conditional on policyholders' attributes, plays an important role in insurance companies' decisions and access risks. Note that there are a plethora of financial/insurance network studies \citep{tang2022insurance} following the collapse and near-failure of the insurance giant American International Group in 2008. Our proposed method analyses the basic risk of the insurance business that is the foundation of these networks. 

This paper proposes a semiparametric estimation method for the conditional joint distribution of bivariate outcomes, using the distribution regression (DR) approach. \cite{williams1972analysis} introduce the DR approach to analyze ordered categorical outcomes by using multiple binary regressions. \citet{foresi1995conditional} first extend the DR approach to characterize a conditional distribution, and various studies \citep[][among others]{hall1999methods,
	chernozhukov2013inference, rothe2013misspecification}
propose the DR approach in different contexts. 

In this paper, we extend the existing research to
the conditional joint distribution of a pair of discrete and continuous outcomes. 
We first apply the factorization formulation of the bivariate joint distribution
and then use the DR method to separately estimate
two conditional distributions:
the distribution of the discrete outcome 
conditional on covariates and
the distribution of the continuous outcome conditional on the covariates and the discrete outcome. Incorporating the discrete outcome as an additional covariate allows us to characterize the dependency between the two outcomes conditional on the covariates in a simple yet flexible manner. Moreover, the combination of the two regression results can uncover the joint conditional distribution and its characteristics.
The joint conditional distribution given a set of covariates can provide a robust statistical basis for computing and optimizing conditional risk measures such as the Value-at-Risk and Expected Shortfall \citep{noyan2013optimization}.

Our approach addresses several issues which may be considered outstanding in multivariate modeling. First, the estimation method in this paper is semiparametric, in that a collection of binary
outcome regressions are used to characterize a joint conditional distribution
through the factorization formulation, instead of imposing global parametric restrictions as of the vast existing literature \citep{olkin1961multivariate,cox1992response,gueorguieva2001correlated}.
Hence, it is useful 
when researchers know little about the underlying distributions
and their parametric form.

Next, our method can be applied to discrete, continuous, or mixed distribution outcomes and flexibly accommodate their dependencies. One of the popular strategies to construct the multivariate distributions conditional on covariates is the copula regression models \citep[see][]{ shi2018pair}. The flexibility of copula models mainly lies in the possibility of specifying the marginals and the dependency among outcome variables separately. The theoretical foundation for the application of copulas is Sklar's theorem, which guarantees the uniqueness of the copula function for continuous outcome variables. In the presence of discrete or mixed distribution outcomes, however,   
the copula function is not unique. To address this issue, \cite{yang2020nonparametric} study nonparametric estimation of copulas for discrete outcomes. For multivariate mixed outcomes, \cite{YangLu2020NCEf} proposes a nonparametric estimator of copulas with the marginal specification based on standard parametric mixed distribution. This specification implies the proportion of zeros in the mixed variables plays a key role in the finite sample performance of the estimator. Our proposed method can serve as an alternative approach in these circumstances. 

Third, our approach is easy to implement and computationally fast among the class of semiparametric and nonparametric methods,
even when the number of covariates is moderately large.
In fact, one can use standard statistical software to implement our approach by fitting only two sets of parametric binary regressions locally over the space of outcome variables. In those local models, parameters can be considered as ``pseudo-parameters", which extract local information of distributions of interest and can be estimated at the parametric rate under certain regularity conditions \citep[see][]{White1982}. We provide the limiting distribution of our estimator, while the limit process depends on unknown nuisance parameters. To circumvent the issue of nonpivotal limit processes, we consider the exchangeable bootstrap  \citep{praestgaard1993Exchangeably} and show its validity, extending the result of \cite{chernozhukov2013inference}.

Our paper complements and extends the recent studies on the multivariate extensions of the DR method. 
\cite{meier2020multivariate} proposes a method to estimate the joint conditional distribution function by directly applying the DR approach with an indicator function over a multi-dimensional grid. As we discussed in Section \ref{sec: Model and Estimation}, this direct application could face a practical issue
even when the number of girds is moderately large.
In contrast, our factorization approach resolves this issue by estimating univariate conditional distribution sequentially. 
Also, \cite{klein2022multivariate} introduce a DR-type approach
that imposes a global structure on the conditional joint distribution function, whereas our approach is semiparametric.

We conduct extensive simulation studies to examine the finite-sample performance of our proposed DR approach under various data generating processes. In particular, we consider two popular parametric models in the insurance literature, the hierarchical model \citep{garrido2016generalized} and the copula model \citep{czado2012mixed},  and a non-standard distribution constructed through transforming a bivariate Gaussian density. The simulation results show that the proposed method performs consistently well across all these setups, whereas the existing parametric approaches perform well only when the model is correctly specified. These results underscore the importance of our semiparametric approach in the context of finite samples.

For empirical application, we analyze a French insurance portfolio. For each policyholder, their characteristics are collected together with their past claim experience, which includes the discrete number of claims made (claim frequency) and the average cost per claim (severity). Traditional parametric approaches have been widely used in the insurance literature \citep{czado2012mixed}, and the increased complexity of insurance data has been driving the development of nonparametric methods. The average cost per claim follows a mixed distribution: a probability mass at zero corresponding to no claims and an otherwise positive claim from a skewed and long-tailed distribution. Hence, a naive specification of parametric claim distributions is often unsatisfactory. The data is very large, consisting of more than 400,000 observations, yet most policyholders did not report any claims. In this case, the copula approach is less robust in capturing the joint distribution  \citep{YangLu2020NCEf}. More importantly, the average severity exhibits a clear multi-modality in the dataset and high skewness when excluding the zero-count observations. This raw data feature suggests that the existing popular parametric approaches are insufficient in this case. The proposed DR approach demonstrates superior performance in both in-sample and out-of-sample results against the existing parametric hierarchical and copula models.

The rest of the paper is organized as follows. In Section 2, we present the DR approach for modeling the bivariate discrete and continuous outcomes. Section 3 sets out the asymptotic properties of our DR method. We provide simulation results in Section 4. In Section 5, we compare our proposed method against existing approaches in application to study a real insurance data set. We conclude this paper in Section 6. The proof of the main results is given in Appendix.

\section{Distribution Regression}
In this section, we illustrate how the DR approach characterizes a univariate conditional distribution.  In the following, we let $W$ be an outcome variable with support $\mathcal{W}\in\mathbb{R}$, which can be discrete, continuous, or mixed, and let $X$ be a $d_x\times1$ vector of covariates with support $\mathcal{X}\in \mathbb{R}^d_x$. The DR approach models the conditional CDF of $W$ given $X=x$ by fitting a parametric linear-index model targeting an arbitrary location of the outcome. More specifically, letting
$\Lambda: \R \to [0,1]$ be a known link function,
we model the conditional distribution function as,
for $(x, w) \in \mathcal{X}{\times}\mathcal{W}$.
\begin{eqnarray}
	\label{eq: uni-DR}
	F_{W|X}(w|x)
	=
	\Lambda\big(P(x)'\alpha(w) \big),
\end{eqnarray}
where $P: \mathcal{X}\rightarrow \mathbb{R}^{d}$ is a known transformation of the conditioning variables, $\alpha(w)\in\mathbb{R}^{d}$ is a vector of unknown parameters specific to the location $w$. The useful link functions include logit, probit, log-log, etc.

Suppose that the data consist of a random sample
$\{(X_{i}, W_{i})\}_{i=1}^{n}$
from the distribution of $(X, W)$
with the sample size of $n$. We can estimate model (\ref{eq: uni-DR}) as binary choice models
for the outcomes $\1\{W \le w\}$ under the maximum likelihood framework, where $\1\{\cdot\}$ is the indicator function:
\begin{eqnarray*}
	\hat{\alpha}(w)
	=
	\arg
	\max_{\alpha}
	\frac{1}{n}
	\sum_{i=1}^{n}\1\{W_{i} \le w \}
	\ln
	\Lambda\big(P(X_i)'\alpha\big)
	+
	\1\{W_{i} > w \}
	\ln
	\big(
	1-
	\Lambda\big(P(X_i)'\alpha\big)
	\big),
\end{eqnarray*}
and then we can estimate the conditional distribution function by
\begin{eqnarray*}
	\widehat{F}_{W|X}(w|x)
	:=
	\Lambda\big(P(x)'\hat{\alpha}(w)\big).
\end{eqnarray*}
Applying the above modeling and estimation procedures on a sequence of locations over the outcome support, the collection of estimation results can characterize the whole conditional distribution.

\section{Model and Estimation}\label{sec: Model and Estimation}
\label{sec:model}

Our interests lie in the conditional distributional features of
bivariate outcomes consisting of continuous and discrete random variables. Practitioners are equipped to grasp the complete picture emerging from the bivariate dependence structure and the influence of specific covariates on various aspects of the variables given the joint conditional distribution. This section outlines the construction of the joint conditional distribution and its associated estimation procedures.

\subsection{Distribution Regression Framework}\label{DR framework}

In what follows, 
we denote by $Y$ a continuous random variable
with the support $\mathcal{Y} \subset \R$
and $Z$ a discrete random variable
with the finite support $\mathcal{Z} \subset \R$.
Let $X$ be a $d_{x} \times 1$ vector of covariates with its support
$\mathcal{X} \subset \R^{d_{x}}$.
We define 
$F_{Y|X,Z}$
and
$F_{Z|X}$
as
the conditional distributions
of $Y$ given $X, Z$
and $Z$ given $X$, respectively.
Then,
we can write 
the joint distribution function
of $(Y, Z)$
conditional on $X$,
using the factorization formulation,
as follows:
for $(x, y, z) \in \mathcal{X}{\times}\mathcal{Y} {\times} \mathcal{Z}$,
\begin{eqnarray}
	\label{eq:joint}
	F_{Y, Z|X}(y, z|x)
	:=
	\int_{\{Z\leq u\}}
	F_{Y|X,Z}(y|x, u)
	d F_{Z|X}(u|x). 
\end{eqnarray}

We estimate the conditional distributions $F_{Y|X,Z}$ and $F_{Z|X}$ by applying the DR method separately and then obtain the joint conditional distribution as in equation (\ref{eq:joint}).
The DR approach fits a parametric linear-index model targeting an arbitrary location of the outcome.
The collection of estimation results over outcome locations can characterize the conditional distribution.
More specifically, letting 
$\Lambda: \R \to [0,1]$ be a known link function,
we model the conditional distribution function as,
for $(x, y, z) \in \mathcal{X}{\times}\mathcal{Y} {\times} \mathcal{Z}$.
\begin{eqnarray} 
	\label{eq:yz}
	F_{Y|X,Z}(y|x,z) 
	=
	\Lambda\big(P_1(x,z)'\alpha(y) \big)
	\ \ \ \ \mathrm{and} \ \ \ \
	F_{Z|X}(z|x)
	=  
	\Lambda\big(P_2(x)'\beta(z) \big),
\end{eqnarray}
where $P_1: \mathcal{X}\times\mathcal{Z}\rightarrow \mathbb{R}^{d_1}$ and $P_2: \mathcal{X}\rightarrow\mathbb{R}^{d_2}$ are two transformations, $\alpha(y)\in\mathbb{R}^{d_1}$ and $\beta(z)\in\mathbb{R}^{d_2}$ are two vectors of unknown parameters.
Those unknown parameters are specific to the points of interest, $y$ or $z$, which can be regarded
as pseudo-parameters to characterize the
conditional distribution at those points,
as discussed in the following subsection.
By setting the link function as the normal or logistic distribution function,
we can consider the models as probit or logit models, respectively.
For each outcome, while one can select a different link function $\Lambda(\cdot)$, we use the same notation for simplicity.

There are several advantages of using the DR method
to estimate the conditional joint distribution in (\ref{eq:joint}).
First, since DR is a local parametric regression,
it is easy to implement and computationally fast,
even when the number of covariates is moderately large.
Second, it characterizes the conditional distributions
by collecting regression results over
the supports $\mathcal{Y}$ and $\mathcal{Z}$.
Thus, the proposed method naturally encapsulates the dependence without global parametric assumptions such as the parametric copula structure. Third, the transformation $P_i$ allows for a flexible enough effect of covariates. For sufficiently rich transformation, one can approximate the conditional distribution function arbitrarily well without extra concern about the choice of the link function.
Lastly, the outcome variable of interest
can be discrete, continuous, or mixed distributions. This extends some existing works of multidimensional distributional regression \citep{klein2022multivariate} generally focusing on continuous distributions. 

\paragraph{Example}
As an illustration, we consider automobile insurance. In an actuarial study, researchers can often observe 
the number of claims $Z$, the average severity $Y$ and
some covariates $X$ for individual policyholders.
Insurance companies face claim losses from each individual policyholder
and the fixed overhead cost of each claim, denoted by $k >0$.
Then,
the aggregate claim amount and the total cost of a policyholder are expressed as
\[
S := Y \cdot Z
\ \ \ \mathrm{and} \ \ \ 
C :=Y \cdot Z +k \cdot Z. 
\] 
The conditional distribution functions
of the aggregate claim
$F_{S|X}(s|x)$ 
and 
of the total cost  $F_{C|X}(c|x)$  
can be written as 
\begin{eqnarray*} 
	F_{S|X}(s|x)
	=
	\left \{
	\begin{array}{ll}
		F_{Z|X}(0|x)
		& \mathrm{if} \ s = 0\\
		F_{Z|X}(0|x)
		+
		\int_{\mathcal{Z} \setminus \{0\}}
		F_{Y|X, Z}
		\big(z^{-1}s |x, z\big)
		d F_{Z|X}(z|x)
		& \mathrm{if} \ s >0
	\end{array}
	\right . ,
\end{eqnarray*} 
and 
\begin{eqnarray*}
	F_{C|X}(c|x)
	=
	F_{Z|X}(0|x)
	+
	\int_{\mathcal{Z} \setminus \{0\}}
	F_{Y|X, Z}
	\big(z^{-1}c - k | x, z\big)
	d F_{Z|X}(z|x).
\end{eqnarray*}
For the purpose of risk management, we can consider a risk measure
as a transformation
of the distribution function to a scalar value. 
For instance,
we can consider 
a Value-at-Risk (VaR) measure
conditional on
policyholders' attributes
$x \in \mathcal{X}$, given by 
\begin{align*}
	VaR_{\tau}(C|x) & :=\inf
	\big \{ c:
	F_{C|X}(c|x) \geq \tau
	\big \}.
\end{align*}
This measure is used to estimate the amount of total cost given
policyholders' information
at a tail event
taking place with
probability $(1-\tau)$. Similarly, we can consider a Expected Shortfall (ES) given by
\begin{align*}
	ES_{\tau}(C|x) & :=\mathbb{E}\big[C|C \geq VaR_{x}(\tau)\big]=\frac{1}{1-\tau}\int_{VaR_{\tau}(C|x)}^{\infty} c\ dF_{C|X}(c|x).
\end{align*}
This measure is used to evaluate the expected loss on a portfolio in the worst $100\tau\%$ of cases.

\subsection{Estimation}

Suppose that the data consist of a random sample 
$\{(X_{i}, Y_{i}, Z_{i})\}_{i=1}^{n}$
from the distribution of $(X, Y, Z)$ 
with the sample size of $n$.
We can consider 
(\ref{eq:yz})
as models that account for 
the probability of the events
$\{Y \le y\}$ and $\{Z \le z\}$
conditional on the covariates.
Thus, 
we can estimate the  models as binary choice models
for the outcomes $\1\{Y \le y\}$ and $\1\{Z \le z\}$
under the maximum likelihood framework.
More specifically, 
the estimators are defined as the maximizers
of the log-likelihood functions,
\begin{eqnarray*}
	\hat{\alpha}(y)
	=
	\arg
	\max_{\alpha}
	\frac{1}{n}
	\sum_{i=1}^{n}
	\ell_{i, y}(\alpha)  
	\ \ \ \mathrm{and} \ \ \ 
	\hat{\beta}(z)
	=
	\arg
	\max_{\beta}
	\frac{1}{n}
	\sum_{i=1}^{n}
	\ell_{i, z}(\beta),
\end{eqnarray*}
where 
\begin{equation}
	\label{eq:lz}
	\begin{aligned}
		\ell_{i, y}(\alpha)
		&:=
		\1\{Y_{i} \le y \}
		\ln
		\Lambda\big(P_1(X_i, Z_i)'\alpha\big)
		+
		\1\{Y_{i} > y \}
		\ln
		\big(
		1-
		\Lambda\big(P_1(X_i, Z_i)'\alpha\big)
		\big),  \\
		\ell_{i, z}(\beta)
		&:=
		\1\{Z_{i} \le z \}
		\ln
		\Lambda\big(P_2(X_{i})'\beta \big)
		+
		\1\{Z_{i} > z \}
		\ln
		\big (
		1- \Lambda\big(P_2(X_{i})'\beta \big)
		\big ). 
	\end{aligned}
\end{equation}
Using the maximum likelihood estimators, we can estimate 
the conditional distributions,
\begin{eqnarray}
	\label{eq:hat}
	\widehat{F}_{Y|X, Z}(y|x, z) 
	:= 
	\Lambda\big(P_1(x,z)'\hat{\alpha}(y)\big)
	\ \ \ \ \mathrm{and} \ \ \ \
	\widehat{F}_{Z|X}(z|x)
	=
	\Lambda\big(P_2(x)'\hat{\beta}(z) \big),
\end{eqnarray}
for any $(x, y, z) \in \mathcal{X}{\times}\mathcal{Y}{\times}\mathcal{Z}$.

\begin{figure}[H]
	\captionsetup[subfigure]{aboveskip=-3pt,belowskip=0pt}
	\centering
	\caption{Implement Illustration for DR Approach}\label{fig: estimation}
	\begin{subfigure}[b]{0.45\textwidth}
		\centering
		\caption{Distribution Estimation of $Z$ given $X=x$}		
		\includegraphics[width=0.95\textwidth]{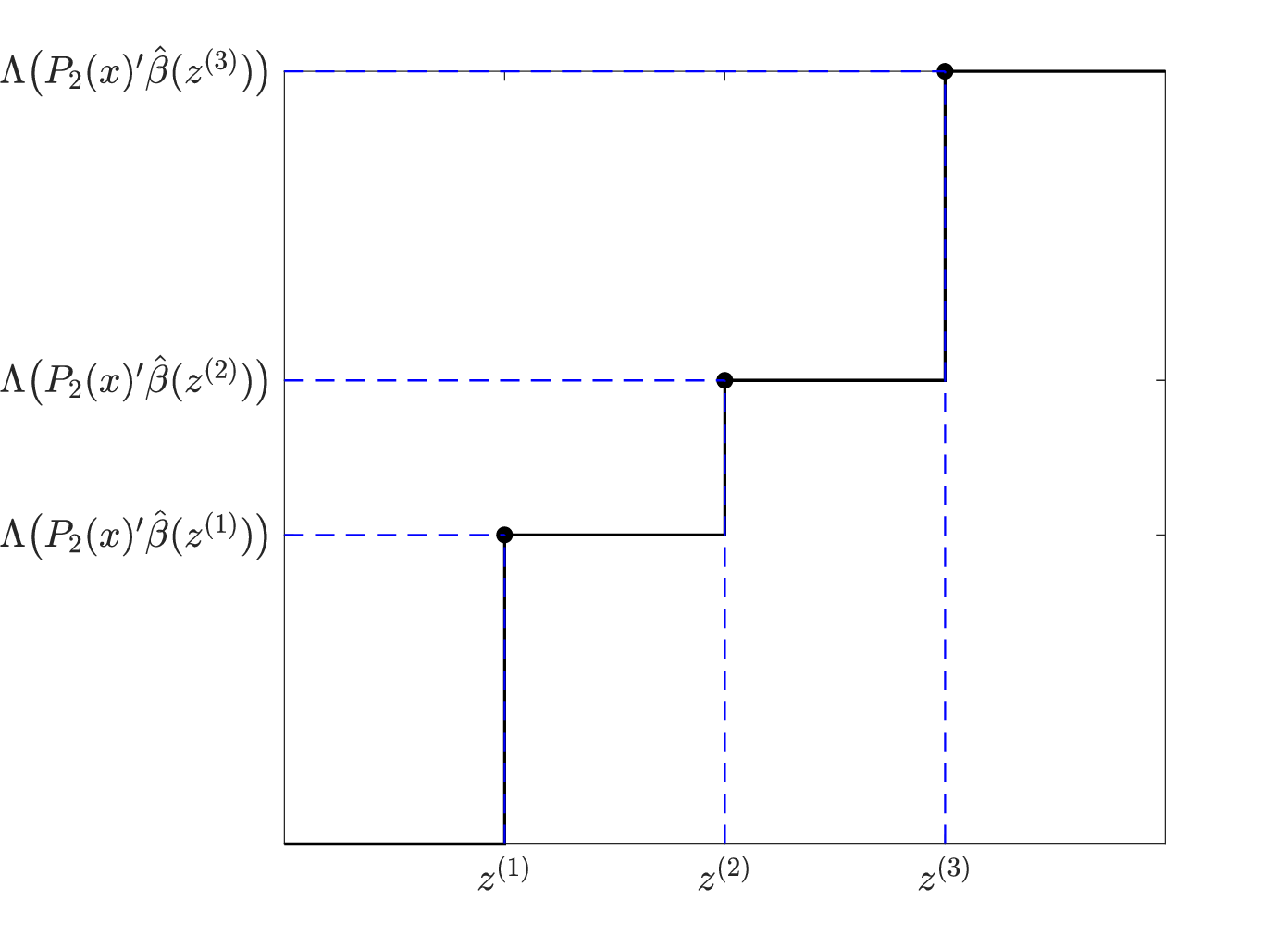}
	\end{subfigure}
    \begin{subfigure}[b]{0.48\textwidth}
     	\centering
     	\caption{Distribution Estimation of $Y$ given $(X, Z) = (x, z)$}     		
     	\includegraphics[width=0.95\textwidth]{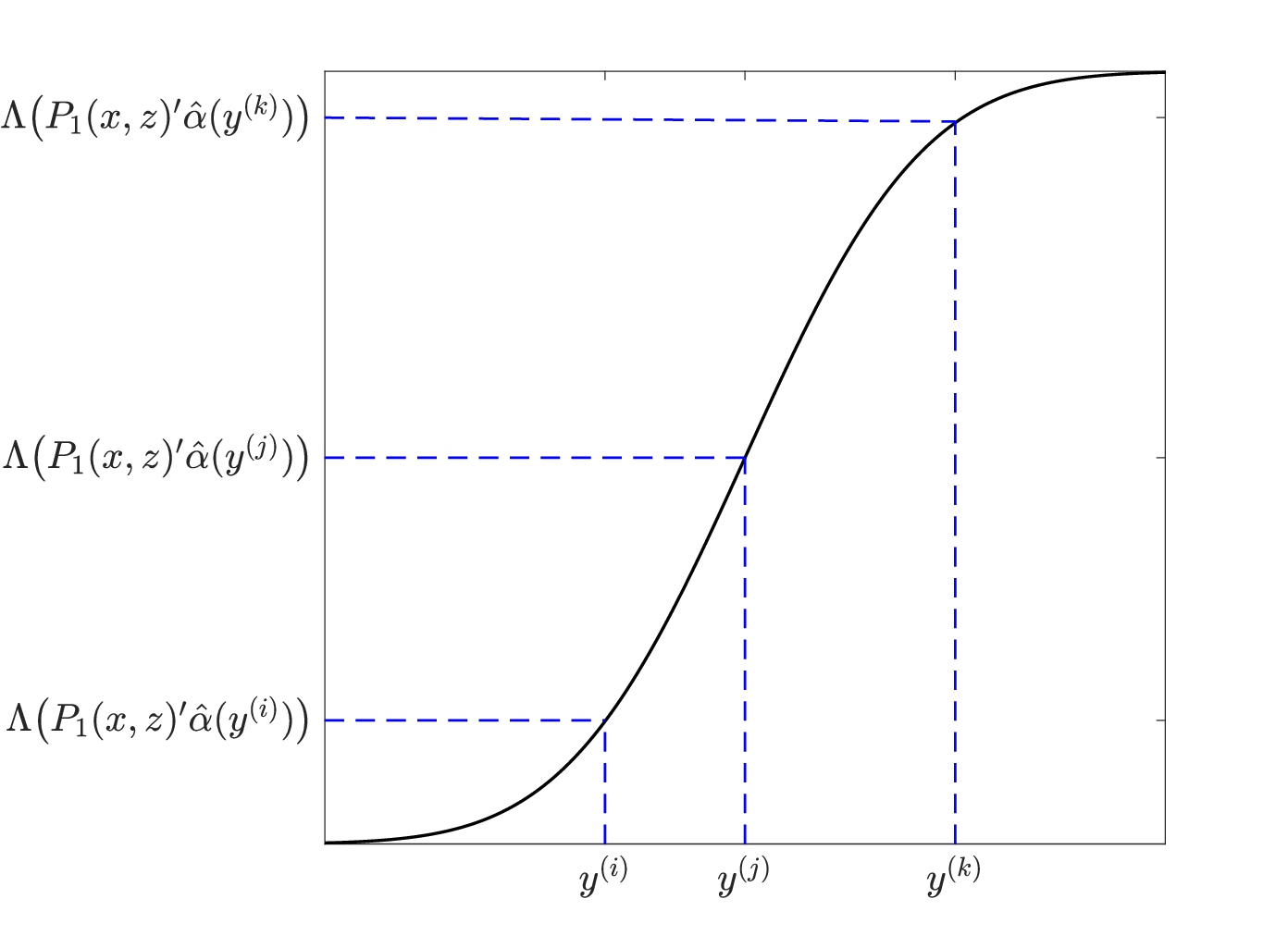}      		
     \end{subfigure}
\end{figure}

In practice, for the discrete variable $Z$ with support $\mathcal{Z}:=\{z^{(1)}, z^{(2)},\ldots, z^{(L)}\}$, we estimate $\widehat{F}_{Z|X}(z^{(l)}|x)$ for all $z^{(l)}\in\mathcal{Z}, l=1,2,\ldots,L$ to construct the estimator $\widehat{F}_{Z|X}$, as illustrated in Figure \ref{fig: estimation}(a). And for the continuous variable $Y$, one can estimate $\widehat{F}_{Y|X, Z}(y^{(j)}|x, z)$ for sufficiently many discrete points $y^{(j)}\in\mathcal{Y}, j=1,2,\ldots,K$ to construct the estimator $\widehat{F}_{Y|X, Z}$, as illustrated in Figure \ref{fig: estimation}(b). Computationally, we are estimating $K+L$ local binary regressions
in total. Our formulation brings computational benefits from the alternative formulation of DR in \cite{meier2020multivariate} building the estimator over a grid which requires $KL$ local optimizations.

One important property that characterizes $\widehat{F}_{Y|X, Z}(y|x, z)$ is monotonicity, i.e., conditional distribution functions
are non-decreasing by definition. Yet, the estimated distribution functions in (\ref{eq:hat}) do not necessarily
satisfy monotonicity in finite samples. We can monotonize the conditional distribution estimators using the rearrangement method proposed by \cite{chernozhukov2009improving}. The rearranging procedure could yield finite-sample improvement
\cite[for instance, see][]{chetverikov2018econometrics} and allow for a straightforward application of the functional delta method
when we transform the estimated distributions by Hadamard differentiable maps.

The introduced estimation procedures can be applied directly to study any variables on a finite support. However, if the variable of interest has an infinite support, such as the interval $[0, \infty)$ or $(-\infty, \infty)$, the DR approach can be augmented with extreme value theory to provide reliable estimates and insights. The extreme value theory offers methodology for studying the tail behavior of the variable, which allows for extrapolation beyond the range of the available data \citep{embrechts2013modelling}. Specifically, to adapt the DR approach for variables with infinite support, we first apply the standard DR approach to the finite support of the variable, which is extracted from the dataset, to construct the conditional distribution. Then, to extrapolate the extreme tail, we fit a generalized extreme value distribution based on several conditional CDF values on the tail of the estimated conditional distribution by the method of moments. Finally, we can obtain the conditional distribution on the whole support by combining the conditional distribution on the finite support with the fitted extreme value distribution on the extreme tails.
\section{Asymptotic Properties and Inference}
\label{sec:Asym}
In this section, we first provide the functional central limit theorems for the estimators of the conditional distribution functions
and their transformations.
Then, we introduce the exchangeable bootstrap for our estimators and establish its validity for practical inference. Detailed proof of all theoretical results is provided in Appendix \ref{sec: appendix-A}.
In what follows,
let
$\|\cdot\|$ be the Euclidean norm for vectors
and
we denote by
$\ell^{\infty}(T)$
the collection of all bounded functions defined on set $T$.

\subsection{Asymptotic Properties}

As the population counterpart of
the log likelihoods,
we define 
$\ell_{y}(\cdot):= \E[\ell_{i, y}(\cdot)]$
and 
$\ell_{z}(\cdot) := \E[\ell_{i, z}(\cdot)]$.
Then, the true parameters
$\alpha_{0}(y)$ and $\beta_{0}(z)$
are defined
as the solution to the following maximization problems,
\begin{eqnarray}
	\label{eq:pl}
	\max_{\alpha}\ell_{y}(\alpha)
	\ \ \ \ \mathrm{and} \ \ \ \
	\max_{\beta}\ell_{z}(\beta).
\end{eqnarray}
We
denote
the second derivative of the population log likelihood
evaluated at the true parameters  
by 
$H_{0, y} := \nabla^2 \ell_{y}\big(\alpha_{0}(y)\big), 
$
and 
$H_{0, z} := \nabla^2 \ell_{z}
\big (
\beta_{0}(z)
\big)$. 
We group the true parameters into a vector 
as
well as
the estimators, by defining 
\begin{eqnarray*}
	\theta_{0}(y, z)
	:= 
	\big[\alpha_{0}(y)', \beta_{0}(z)'\big]'
	\ \ \ \mathrm{and} \ \ \
	\hat{\theta}(y, z)
	:= 
	\big[\hat{\alpha}(y)', \hat{\beta}(z)'\big]',
\end{eqnarray*}
and let $\Theta$ denote the parameter space.\footnote{
	The parameter space can be defined for
	each $(y, z) \in \mathcal{Y}{\times}\mathcal{Z}$,
	while we suppress the dependency
	for notational simplicity.
}
Also, for $\theta:=(\alpha', \beta')\in \Theta$, we introduce
a vector of the first derivatives
of functions 
in (\ref{eq:lz}) as 
\begin{eqnarray*}
	\varphi_{\theta, y, z}(X, Y, Z)
	:=
	\left(
	\begin{array}{c}
		\big[\1\{ Y\leq y\}-\Lambda \big(P_1(X,Z)'\alpha \big)\big]
		R\big(P_1(X,Z)'\alpha \big)P_1(X,Z) \\
		\big[\1\{ Z\leq z\}-\Lambda \big(P_2(X)'\beta \big)\big]
		R\big(P_2(X)'\beta \big)P_2(X) 
	\end{array}
	\right),
\end{eqnarray*}
where $R(u):=\lambda(u)/\big\{\Lambda(u)[1-\Lambda(u)]\big\}$.

To obtain the asymptotic
results, the following assumptions are imposed.

\vspace{0.5cm}
\noindent  
\textbf{Assumptions}:
\begin{itemize}
	\item[A1.]
	The observations 
	$\big\{(X_{i}, Y_{i}, Z_{i}) \in \mathcal{X}{\times}\mathcal{Y} {\times} \mathcal{Z}\big\}_{i=1}^{n}$
	are
	independent and identically distributed (iid).
	The supports
	$\mathcal{X}$
	and
	$\mathcal{Y}$
	are compact 
	and
	$\mathcal{Z}$
	is a finite set of discrete points.
	
	\item[A2.]
	For any $y \in \mathcal{Y}$ and $z \in \mathcal{Z}$,
	the log-likelihood functions 
	$\alpha\mapsto
	n^{-1} \sum_{i=1}^{n}
	\ell_{i, y}(\alpha)$
	and
	$\beta \mapsto
	n^{-1} \sum_{i=1}^{n}
	\ell_{i,z}(\beta)$
	are concave 
	for their arguments.
	The link function $\Lambda(\cdot)$
	is 
	twice continuously differentiable
	with its first derivative $\lambda(\cdot)$.
	
	\item[A3.]
	The true parameters
	$\theta_{0}(y,z)$
	uniquely
	solve 
	the maximization problem in (\ref{eq:pl})
	and
	are contained in
	the interior
	of 
	the compact parameter space $\Theta$.
	
	\item[A4.]
	The maximum eigenvalues of 
	$H_{0, y}$
	and
	$H_{0, z}$
	are strictly negative
	uniformly
	over $y \in \mathcal{Y}$
	and $z \in \mathcal{Z}$. 
	
	\item[A5.]  
	The conditional density function  
	$f_{Y|X, Z}(y|x,z)$ exists, 
	is uniformly bounded in $(y, x) \in \mathcal{Y} \times \mathcal{X}$, 
	and is uniformly continuous in $y\in\mathcal{Y}$ for any $x\in\mathcal{X}$.
\end{itemize}
\vspace{0.5cm}


Assumption A1 is imposed in the research conducted by \cite{chernozhukov2013inference}. The bounded supports of covariates ensures that $\E\|X_{i}\|^2 < \infty$. It's worth noting that the compactness assumption on the supports is essential for the uniform valid statistical inference over the entire state space, but it is not necessary for estimation purposes.
Assumption A2 ensures that
standard optimization procedures based on derivatives  
can easily obtain the maximum likelihood estimators.
A similar condition is assumed in the research by \cite{chernozhukov2013inference}
and 
both the logit and probit links satisfy this condition.
Assumption A3 guarantees the existence of the true parameters.
Even when the model in (\ref{eq:yz}) is miss-specified,
we can consider the true parameters
as pseudo-parameters satisfying the first-order conditions,
$\nabla \ell_{y}
\big (
\alpha_{0}(y)
\big) =0$
and 
$\nabla \ell_{z}
\big (
\beta_{0}(z)
\big) = 0$,
under assumptions A2 and A3,
and
thus
the estimators of the parameters 
can be interpreted under the quasi-likelihood framework 
for each $y \in \mathcal{Y}$ and $z \in \mathcal{Z}$
\citep[see][]{Huber1967, White1982}.  
Assumption A4 is required to ensure that the information matrices are invertible
over the supports. 
Assumption A5 is required to obtain 
the limit process of our estimators over the supports for statistical inference. 

Under the assumptions above,
the proposition below provides
the limit process of
the estimators
$
\hat{\theta}(y, z)
$
over $\mathcal{Y} \times \mathcal{Z}$.

\vspace{0.5cm}
\begin{proposition}
	\label{pro:est}
	Suppose that Assumptions A1-A5 hold.
	Then, we have 
	\begin{eqnarray*}
		\sqrt{n}
		\big (
		\hat{\theta}(\cdot, \cdot)
		-
		\theta_{0}(\cdot, \cdot)
		\big )
		\rightsquigarrow
		\mathbb{B}(\cdot,\cdot)
		\ \ \mathrm{in} \ \
		\ell^{\infty}(\mathcal{Y})^{d_1}
		{\times}
		\ell^{\infty}(\mathcal{Z})^{d_2},
	\end{eqnarray*}
	where
	$\mathbb{B}(y,z)$ is
	a mean-zero Gaussian process
	over 
	$\mathcal{Y} \times \mathcal{Z}$,
	and
	its covariance function is
	given by
	$
	H_{0}(y_{1},z_{1})^{-1}
	\Sigma(y_{1}, z_{1}, y_{2}, z_{2})
	H_{0}(y_{2},z_{2})^{-1} 
	$
	for $(y_{1},z_{1}), (y_{2},z_{2}) \in \mathcal{Y} \times \mathcal{Z}$,
	with
	a block diagonal matrix 
	$H_{0}(y, z):=\mathrm{diag}
	(
	H_{0, y},
	H_{0, z}
	)$
	and 
	$
	\Sigma(y_{1}, z_{1}, y_{2}, z_{2}):=
	\E[\varphi_{\theta_{0}(y_{1},z_{1}), y_{1},z_{1}} \varphi_{\theta_{0}(y_{2},z_{2}), y_{2},z_{2}}']$.
\end{proposition}
\vspace{0.5cm}

In the proof of Proposition \ref{pro:est},
we use the convexity property in Assumption A2 to obtain
the limiting processes over the support $\mathcal{Y}$, 
following the argument used for quantile regression \citep[see][]{pollard1991asymptotics, kato2009asymptotics}.
The result in Proposition \ref{pro:est} shows that the covariance function exhibits the sandwich form as the covariance matrix is obtained under the quasi-likelihood framework.

The distribution function estimators, $\widehat{F}_{Y|X, Z}$
and $\widehat{F}_{Z|X}$, are a transformation of the estimator $\hat{\theta}(\cdot)$ as in (\ref{eq:hat}).
Let 
$\mathbb{D}:=
\ell^{\infty}(\mathcal{Y})^{d_1}
{\times}
\ell^{\infty}(\mathcal{Z})^{d_2} $
and 
define 
$
\phi( \hat{\theta})
=
\big (
\widehat{F}_{Y|X, Z}, 
\widehat{F}_{Z|X}
\big )
$,
where 
the map 
$\phi: \mathbb{D}_{\phi}\subset \mathbb{D}\mapsto \mathbb{S}_{\phi}$,
given by
\begin{eqnarray*}
	\phi(\theta)(x, y, z) :=
	\big[\Lambda\big(P_1(x,z)'\alpha(y)\big),
	\Lambda\big(P_2(x)'\beta(z)\big)
	\big]'.
\end{eqnarray*}
It can be shown that 
the map $\phi$
is 
Hadamard differentiable
at $\theta \in \mathbb{D}_{\phi}$
tangentially to $\mathbb{D}$
with
the Hadamard derivative 
$(a, b)
\mapsto 
\phi_{\theta_{0}(\cdot)}'(a, b)$
is 
given by 
\begin{eqnarray*}
	\phi_{\theta_{0}(\cdot)}'(a, b)(x,y,z)
	:=
	\big [
	\lambda
	\big (P_1(x,z)'\alpha_{0}(y)\big)
	P_1(x,z)'a(y), 
	\lambda
	\big (
	P_2(x)'\beta_{0}(z)
	\big)
	P_2(x)' b(z)
	\big] '.
\end{eqnarray*}
The theorem below shows
the joint asymptotic distribution of the distribution function
estimators, applying the functional delta method
with the Hadamard derivative in the above display.
Furthermore, 
we can easily derive the asymptotic distribution
of the estimator of distributional characteristics,
such as the  Value-at-Risk conditional on covariates
and distributional features of $Y$ and $Z$ after some transformation
if the distributional characteristics are obtained 
through Hadamard differentiable maps.

\vspace{0.5cm}
\begin{theorem}
	\label{theorem:CLT}
	Suppose that Assumptions A1-A5 hold.
	Then,
	\begin{itemize}
		\item [(a)]
		we have 
		\begin{eqnarray*}
			\sqrt{n}
			\left (
			\begin{array}{c}
				\widehat{F}_{Y|X, Z}
				-
				F_{Y|X, Z} \\
				\widehat{F}_{Z|X}
				-
				F_{Z|X}
			\end{array}
			\right )
			\rightsquigarrow
			\phi_{\theta_{0}(\cdot)}'
			\big (
			\mathbb{B}
			\big )
			\ \ \mathrm{in} \ \
			\ell^{\infty}(\mathcal{X}{\times}\mathcal{Y}{\times}\mathcal{Z})
			{\times}
			\ell^{\infty}(\mathcal{X}{\times}\mathcal{Z}),
		\end{eqnarray*}
		where
		$\mathbb{B}$ is the mean-zero Gaussian process defined in Proposition \ref{pro:est};
		\item[(b)]
		additionally, 
		if a map 
		$\nu:
		\mathbb{S}_{\phi} 
		\to
		\ell^{\infty}(\mathcal{X}{\times}\mathcal{Y}{\times}\mathcal{Z})
		$
		is Hadamard differentiable
		at
		$(F_{Y|X,Z}, F_{Z|X})$
		tangentially to
		$\phi_{\theta}'(\mathbb{D})$ 
		with
		the derivative
		$\nu_{F_{Y|X,Z}, F_{Z|X}}'$,
		then 
		\begin{eqnarray*}
			\sqrt{n}
			\big \{
			\nu
			\big(
			\widehat{F}_{Y|X, Z},
			\widehat{F}_{Z|X}
			\big) 
			-
			\nu(F_{Y|X,Z}, F_{Z|X}) 
			\big \}
			\rightsquigarrow
			\nu_{F_{Y|X,Z}, F_{Z|X}}'
			\circ
			\phi_{\theta_{0}(\cdot)}'
			\big (
			\mathbb{B}
			\big ),
		\end{eqnarray*}
		in
		$  \ell^{\infty}(\mathcal{X}{\times}\mathcal{Y}{\times}\mathcal{Z})$.
	\end{itemize}
	
\end{theorem}
\vspace{0.5cm}

The limiting processes presented in
the above proposition and theorem
depend on 
unknown nuisance parameters
and may complicate inference in finite samples.
The subsequent subsection 
introduces the
bootstrap scheme
and reveals its validity.

\subsection{Exchangeable Bootstrap}

To deal with the issue of nonpivotal limit processes, 
we consider a resampling method called the exchangeable bootstrap  
\citep[see][]{praestgaard1993Exchangeably, van1996weak}.  
This resampling scheme consistently estimates  
limit laws of relevant empirical distributions
and thus, using the functional delta method, 
consistently estimates the limit process of the estimator.

For the resampling scheme, we introduce a vector of random weights 
$(W_{1}, \dots, W_{n})$.
To establish the validity of the bootstrap, 
we assume that the random weights satisfy the 
following conditions.

\vspace{0.5cm}
\noindent 
\textbf{Assumption B}. 
Let
$(W_{1}, \dots, W_{n})$
be $n$ scalar, nonnegative  
random variables,
which 
are identically distributed, independent of the original sample,
and
satisfy the following conditions: for some $\epsilon>0$,
\begin{eqnarray*}
	\E|W_{1}|^{2+\epsilon}
	< \infty,
	\ \ 
	\bar{W}_{n}
	:=
  \frac{1}{n}
	\sum_{i=1}^{n}
	W_{i}
	\to^p
	1,
	\ \
  \frac{1}{n}
	 \sum_{i=1}^{n}
	(
	W_{i}
	- 
	\bar{W}_{n}
	)^2
	\to^p 1.
\end{eqnarray*}
\vspace{0.2cm}

As \cite{van1996weak} explain, this resampling scheme encompasses 
a variety of bootstrap methods, such as 
the empirical bootstrap, subsampling, wild bootstrap and so on. 
These conditions are employed by \cite{chernozhukov2013inference}
for the inference of counterfactual distributions.

Given the random weights, 
we obtain 
the bootstrap  estimator
\begin{eqnarray*}
	\hat{\theta}^{*}(y,z):=
	\big(
	\hat{\alpha}^{*}(y)',
	\hat{\beta}^{*}(z)'
	\big)'  
\end{eqnarray*}
by maximizing the log likelihoods:
\begin{eqnarray*}
	\hat{\alpha}^{\ast}(y)
	=
	\arg
	\max_{\alpha}
	\frac{1}{n}
	\sum_{i=1}^{n}
	W_{i}
	\cdot
	\ell_{i, y}(\alpha)
	\ \ \ \mathrm{and} \ \ \ 
	\hat{\beta}^{\ast}(z)
	=
	\arg
	\max_{\beta}
	\frac{1}{n}
	\sum_{i=1}^{n}
	W_{i}
	\cdot
	\ell_{i, z}(\beta). 
\end{eqnarray*}
Then, we can obtain the bootstrap counterparts of
the conditional distribution estimators:
\begin{eqnarray*}
	\widehat{F}_{Y|X, Z}^{\ast}(y|x, z) 
	:=
	\Lambda\big(P_1(x,z)' \hat{\alpha}^{\ast}(y) \big)
	\ \ \ \ \mathrm{and} \ \ \ \
	\widehat{F}_{Z|X}^{\ast}(z|x)
	:=
	\Lambda\big(P_2(x)'\hat{\beta}^{\ast}(z) \big), 
\end{eqnarray*}
as well as their transformation
$    \nu
\big(
\widehat{F}_{Y|X, Z}^{\ast},
\widehat{F}_{Z|X}^{\ast}
\big) $.

For the validity of the resampling method explained above, we need to introduce the notion
of conditional weak convergence in probability,
following \cite{van1996weak}.
For some normed space $\mathbb{Q}$,
let $BL_{1}(\mathbb{Q})$
denote the space of all 
Lipschitz continuous functions
from $\mathbb{Q}$ to $[-1,1]$.
Given the original sample
$\{(X_{i}, Y_{i}, Z_{i})\}_{i=1}^{n}$,
consider a random element 
$B_{n}^{\ast}:=
g(\{(X_{i}, Y_{i}, Z_{i})\}_{i=1}^{n}, \{W_{i}\}_{i=1}^{n})$
as  
a function of 
the original sample 
and 
the random weight vector 
generating 
the bootstrap draw. 
The bootstrap law of $B_{n}^{\ast}$
is said to 
consistently estimate the law of some tight random element 
$B$ or 
$B_{n} \rightsquigarrow^{p} B$ 
if 
\begin{eqnarray*}
	\sup_{h \in BL_{1}(\mathbb{Q})}
	\big |
	\E_{n}[h(B_{n}^{\ast})]
	-
	\E[h(B)]
	\big |
	\to^p 0,
\end{eqnarray*}
where 
$\E_{n}$ is the expectation with respect to
$\{W_{i}\}_{i=1}^{n}$
conditional on the original sample.

In the theorem provided below, we first show that 
the exchangeable bootstrap provides a method to consistently estimate 
the limit process of a pair of conditional distributions.
Additionally, we show that the limit process of the Hadamard differentiable
transform can be estimated using the functional delta method.

\vspace{0.5cm}
\begin{theorem}
	\label{theorem:bootstrap}  
	Suppose that Assumptions A1-A5 and B hold. Then,
	\begin{itemize}
		\item[(a)] we have 
		\begin{eqnarray*}
			\sqrt{n}
			\left (
			\begin{array}{c}
				\widehat{F}_{Y|X, Z}^{\ast}
				-
				\widehat{F}_{Y|X, Z} \\
				\widehat{F}_{Z|X}^{\ast}
				-
				\widehat{F}_{Z|X}
			\end{array}
			\right )
			\rightsquigarrow^{p}
			\phi_{\theta_{0}(\cdot)}'
			\big (
			\mathbb{B}
			\big )
			\ \ \mathrm{in} \ \
			\ell^{\infty}(\mathcal{X}{\times}\mathcal{Y}{\times}\mathcal{Z})
			{\times}
			\ell^{\infty}(\mathcal{X}{\times}\mathcal{Z}),
		\end{eqnarray*}
		\item[(b)] 
		additionally, 
		if the map 
		$\nu:
		\mathbb{S}_{\phi} 
		\to
		\ell^{\infty}(\mathcal{X}{\times}\mathcal{Y}{\times}\mathcal{Z})
		$
		is Hadamard differentiable
		at
		$(F_{Y|X,Z}, F_{Z|X})$
		tangentially to
		$\phi_{\theta}'(\mathbb{D})$ 
		with
		the derivative
		$\nu_{F_{Y|X,Z}, F_{Z|X}}'$,
		then
		\begin{eqnarray*} 
			\sqrt{n}
			\big \{
			\nu
			\big(
			\widehat{F}_{Y|X, Z}^{\ast},
			\widehat{F}_{Z|X}^{\ast}
			\big) 
			-
			\nu(
			\widehat{F}_{Y|X, Z},
			\widehat{F}_{Z|X}) 
			\big \}
			\rightsquigarrow^{p}
			\nu_{F_{Y|X,Z}, F_{Z|X}}'
			\circ
			\phi_{\theta_{0}(\cdot)}'
			\big (
			\mathbb{B}
			\big ),
		\end{eqnarray*}
		in
		$  \ell^{\infty}(\mathcal{X}{\times}\mathcal{Y}{\times}\mathcal{Z}) $.
	\end{itemize}
	
\end{theorem}
\vspace{0.5cm}
In practice,
we monotonize the bootstrap counterparts of the conditional distribution estimators, using the rearrangement method proposed by \cite{chernozhukov2009improving}.

\section{Monte-Carlo Simulations}
\label{sec:simulation}  
This section presents Monte Carlo
simulation results to reveal the finite-sample properties of the proposed method,
compared with the existing methods.

\subsection{Simulation Setup}

 Let $F_{Z}$ and $F_{Y}$ be two parametric distributions left to be specified for $Z$ and $Y$, respectively, and let $g_{Z}$ and $g_{Y}$ be two proper link functions for modeling the conditional means of $Z$ and $Y$ under GLM framework, respectively. We compare our method with the following two popular parametric models: the hierarchical model of \cite{garrido2016generalized} 
\begin{equation}
	\begin{array}{ll}
		Z|X\sim F_{Z}, & \mu_Z:=\mathbb{E}(Z|X)=g_Z(X'\beta),\\
		Y|X,Z\sim F_{Y}, & \mu_Y:=\mathbb{E}(Y|X,Z)=g_Y(Z\gamma+X'\alpha),		
	\end{array}
	\label{eq:PG}
\end{equation}
and the Gaussian copula regression model of \cite{czado2012mixed} 
\begin{equation}
	\begin{array}{l}
		\begin{array}{ll}
			Z|X\sim F_{Z}, & \mu_Z:=\mathbb{E}(Z|X)=g_Z(X'\beta),\\
			Y|X\sim F_{Y}, & \mu_Y:=\mathbb{E}(Y|X)=g_Y(X'\alpha),			
		\end{array} \\
		F_{Y,Z|X}(y,z|x)=C_{\eta}
		\big (F_{Y}(y|x),F_{Z}(z|x) \big),
	\end{array}
	\label{eq:copula}
\end{equation}
where $C_{\eta}:[0,1]^{2} \to [0,1]$
is the Gaussian copula with the correlation parameter $\eta$.

We consider the three different data generating processes (DGPs) below for
obtaining the samples of $\big\{(X_{i}, Y_{i}, Z_{i})\big\}_{i=1}^{n}$
with
the sample size
$n=2,000$. 
We set $\alpha=(0.5,1,1,1)'$ for all DGPs,
while we consider two cases of $\beta$ for each DGP
in order to consider
that 
the probability of $Z$ taking zero differs across the cases,
as follows:
\begin{enumerate}
	\item[] DGP 1. Hierarchical model in (\ref{eq:PG}) with
	\[F_{Z}\sim Poisson(\mu_Z),\ \ \ \ F_{Y}\sim GB2(\mu_Y, \sigma, k_1, k_2)\footnote{The density of $GB2(\mu_Y,\sigma,k_1,k_2)$ is $g(y)=\frac{\exp \left(k_1 z\right)}{y \sigma B\left(k_1, k_2\right)[1+\exp (z)]^{k_1+k_2}}$ with $z=(\log y-mu_Y)/\sigma$.}, \]
	and $(\gamma, \sigma, k_1, k_2) =(-0.5, 0.5, 5, 3.5)$;
	\begin{enumerate}
		\item[] Case 1:
		$\beta=(0.5,-0.5,-0.5,-0.5)'$,
		\ and \
		Case 2:
		$\beta=(-1,0.5,0.5,0.5)'$.
	\end{enumerate}
	
	\item[] DGP 2.  Gaussian copula model in (\ref{eq:copula}) with 
		\[F_{Z}\sim Poisson(\mu_Z),\ \ \ \ F_{Y}\sim Gamma(\mu_Y, \delta), \]
	and $(\delta, \eta)=(0.2, -0.5)$;
	\begin{enumerate}
		\item[]Case 1:
		$\beta=(-1,1,1,1)'$,
		\ and \ Case 2: $\beta=(-2,0.6,0.6,0.6)'$.
	\end{enumerate}
	
	\item[] DGP 3. Truncated Bivariate Normal DGP defined as:
	\begin{eqnarray*}
		Y = |S_{1}|
		\ \ \mathrm{and} \ \  Z=z \ \mathrm{if} \ z- 1 < |S_{2}|\leq z
	\end{eqnarray*}
	for non-negative integer $z$,
	where 
	$
	(S_{1},S_{2})
	$
	is bivariate normally distributed
	conditional on $X$
	with 
	$\E\big[(S_{1}, S_{2})\big]=
	\left(
	X'\alpha,
	X'\beta\right)
	$,
	$\mathrm{var}(S_{1})=1$,
	$\mathrm{var}(S_{2})=40$
	and
	$\mathrm{cov}(S_{1}, S_{2})=5$;
	\begin{enumerate}
		\item[]Case 1: $\beta=(1.5,0.2,0.2,0.2)'$,
		\ and \
		Case 2: $ \beta=(0.1,0.2,0.2,0.2)'.$
	\end{enumerate}
\end{enumerate}
We select parameter values to ensure that
$\Pr(Z=0)$ is roughly between 0.20-0.25
and 0.65-0.70
under Case 1 and Case 2, respectively.
In all DGPs, we assume that $Y$ equals 0 if $Z$ is 0, thus the outcome $Y$ follows a mixed distribution.
We use regressors $X=\big(1, X^{(1)},X^{(2)},X^{(3)}\big)'$ with $X^{(j)}$ randomly generated from the standard uniform distribution for $j=1, 2, 3$.

For all estimation models, we first model the two univariate conditional distributions of $Z$ and $Y|Y>0$, given $P(Y=0|X)\equiv P(Z=0|X)$, the mixed distribution of $Y$ and the joint distribution of $(Y,Z)$ can be obtained. For comparisons under DGPs 1 and 2, the true marginal distribution families are assumed for $F_Z$ and $F_Y$ of the two parametric models. Thus, the hierarchical model in (\ref{eq:PG}) and the copula model in (\ref{eq:copula}) are correctly specified under DGPs 1 and 2, respectively.  On the other hand, the DR is miss-specified in all DGPs. Taking miss-specification into account, we consider a transformation that includes pairwise products of regressors in addition to the original regressors for all estimation models. For the DR approach, the logit link function is applied and the discretization points of the support of $Y$ are chosen as empirical quantiles of $\{Y_{i}\}_{i=1}^{n}$ for probabilities 1\%, 2\%, $\ldots$, 100\%.

\subsection{Simulation Results}
We compare the performance of our DR method with the two competing models on estimating the conditional mean, conditional standard deviation, 95\% ES, and 0.95th conditional quantiles of $C=Y\cdot Z+Z$ given $X=x$, which are denoted by $\mathbb{E}(C|x)$, $Std(C|x)$, $ES_{0.95}(C|x)$ and $Q_{0.95}(C|x)$, respectively. For the DR method, we obtain the conditional CDF $F_{C|X}(c|x)$ across a series of discrete points over the support of $C$, which allows us to estimate these measures numerically. Specifically, upon obtaining the sequence of conditional CDF values, we could employ the inverse transformation method to generate a $10,000$ random samples, and approximate the desired measures by their sample counterparts.
We consider values of regressors $x = (1, x_{1}, 0.5, 0.5)'$
with $x_1$ taking 0.25, 0.50, or 0.75 for comparison.
Under each DGP, we present the estimated errors measured by bias and mean square of errors (MSE) based on 1,000 times Monte Carlo simulations.

\paragraph{Poisson-GB2 Hierarchical DGP}

For this DGP, we assume that $F_Z$ and $F_Y$ follow the Poisson and GB2 distributions, respectively, for both parametric models. Accordingly, the log and identical link functions are adopted for $g_Z$ and $g_Y$, respectively. Thus, under this DGP, the hierarchical model is correctly specified, and the Gaussian copula regression uses the correctly specified margins but the misspecified dependence structure.

\begin{table}[H] 
	\centering
	\caption{Monte Carlo Results under Poisson-GB2 Hierarchical DGP\label{tab: PGB2-C}}
	\linespread{1.0}
	\small
	\begin{tabular}{llrlrrrrrrr}
		\hline
		\hline
		&                         & \multicolumn{1}{l}{} &  & \multicolumn{3}{c}{Bias}                                              & \multicolumn{1}{c}{} & \multicolumn{3}{c}{MSE}                                               \\ \cline{5-7} \cline{9-11} 
		Quantities                         & Case                    & $x_1$                &  & \multicolumn{1}{c}{DR} & \multicolumn{1}{c}{H} & \multicolumn{1}{c}{Copula} & \multicolumn{1}{c}{} & \multicolumn{1}{c}{DR} & \multicolumn{1}{c}{H} & \multicolumn{1}{c}{Copula} \\ \hline
		\multirow{6}{*}{$\mathbb{E}(C|x)$} &{Case 1} & 0.25                 &  & 0.08                  & 0.04                  & 0.56                  &                      & 0.06                  & 0.03                  & 0.36                  \\
		&                         & 0.50                 &  & 0.05                  & 0.01                  & 0.55                  &                      & 0.09                  & 0.04                  & 0.38                  \\
		&                         & 0.75                 &  & 0.06                  & 0.02                  & 0.58                  &                      & 0.08                  & 0.05                  & 0.41                  \\
		&{Case 2} & 0.25                 &  & -0.01                 & -0.03                 & 0.32                  &                      & 0.03                  & 0.02                  & 0.14                  \\
		&                         & 0.50                 &  & -0.003                & -0.03                 & 0.52                  &                      & 0.07                  & 0.04                  & 0.33                  \\
		&                         & 0.75                 &  & 0.09                  & 0.04                  & 0.86                  &                      & 0.15                  & 0.06                  & 0.85                  \\ \hline
		\multirow{6}{*}{$Std(C|x)$}        & {Case 1} & 0.25                 &  & 0.024                 & 0.002                 & 1.57                  &                      & 0.01                  & 0.01                  & 2.51                  \\
		&                         & 0.50                 &  & -0.008                & -0.033                & 1.71                  &                      & 0.03                  & 0.01                  & 2.98                  \\
		&                         & 0.75                 &  & 0.020                 & -0.004                & 1.93                  &                      & 0.04                  & 0.02                  & 3.79                  \\
		& {Case 2} & 0.25                 &  & 0.002                 & -0.014                & 1.19                  &                      & 0.01                  & 0.01                  & 1.44                  \\
		&                         & 0.50                 &  & -0.040                & -0.047                & 1.71                  &                      & 0.02                  & 0.01                  & 2.98                  \\
		&                         & 0.75                 &  & 0.001                 & -0.001                & 2.52                  &                      & 0.04                  & 0.02                  & 6.44                  \\ \hline
		\multirow{6}{*}{$ES_{0.95}(C|x)$}     &  {Case1}  & 0.25                     &                      & 0.11                   & -0.02                   & 7.87                  &                      & 0.22                   & 0.14                   & 62.51                    \\
		&                        & 0.5                      &                      & -0.10                 & -0.20                   & 8.70                   &                      & 0.42                   & 0.27                    & 76.58                    \\
		&                        & 0.75                     &                      & -0.09                   & -0.10                    & 9.88                    &                      & 0.82                  & 0.35                   & 98.86                   \\
		&  {Case2}  &  0.25 &  & 0.10  & 0.01  & 6.42  &  & 0.18 & 0.12 & 41.75  \\
		&             & 0.5  &  & -0.18 & -0.23 & 8.72  &  & 0.34 & 0.27 & 76.91  \\
		&             & 0.75 &  & 0.08  & 0.05  & 12.34 &  & 0.69 & 0.39 & 153.72 \\ \hline		
		\multirow{6}{*}{$Q_{C}(0.95|x)$}   & {Case 1} & 0.25                 &  & -0.06                 & 0.01                  & 4.66                  &                      & 0.10                  & 0.06                  & 22.02                 \\
		&                         & 0.50                 &  & -0.13                 & -0.05                 & 5.01                  &                      & 0.20                  & 0.10                  & 25.58                 \\
		&                         & 0.75                 &  & 0.03                  & 0.09                  & 5.51                  &                      & 0.39                  & 0.15                  & 31.04                 \\
		& {Case 2} & 0.25                 &  & -0.13                 & -0.05                 & 3.43                  &                      & 0.10                  & 0.06                  & 12.03                 \\
		&                         & 0.50                 &  & -0.33                 & -0.20                 & 4.89                  &                      & 0.27                  & 0.13                  & 24.35                 \\
		&                         & 0.75                 &  & -0.03                 & 0.08                  & 7.34                  &                      & 0.32                  & 0.16                  & 54.68                 \\ \hline
	\end{tabular}
	\begin{minipage}{0.85\linewidth} \linespread{1.0}\footnotesize
		\textit{Notes}:
		The number of
		Monte Carlo iterations
		is set to 1,000.
		We choose covariates $x=(1, x_1,0.5,0.5)$ with $x_1\in\{0.25, 0.50, 0.75\}$.
		For each quantity, we report the bias and MSE in both cases. For simplicity, we represent the hierarchical and Gaussian copula models as `H' and `Copula', respectively.
	\end{minipage}
\end{table} 

All of the simulation results are presented in Table \ref{tab: PGB2-C}. As expected, the hierarchical model has the best estimation performance, while the DR approach can always provide a comparative performance in most cases. In particular, the DR approach performs better than the hierarchical model on estimating the ES. Both of them consistently outperform the Gaussian copula regression model for all quantities.

\paragraph{Poisson-Gamma Gaussian Copula DGP}

\begin{table}[H]
	\centering
	\caption{ Monte Carlo Results under Poisson-Gamma Gaussian Copula DGP\label{tab: Copula-C}}
	\linespread{1.0}
	\small
	\begin{tabular}{llrrrrrrrrr}
		\hline
		\hline
		&                         & \multicolumn{1}{l}{} &  & \multicolumn{3}{c}{Bias}                                                   & \multicolumn{1}{c}{} & \multicolumn{3}{c}{MSE}                                                    \\ \cline{5-7} \cline{9-11} 
		Quantities                         & Case                    & $x_1$                &  & \multicolumn{1}{c}{DR} & \multicolumn{1}{c}{H} & \multicolumn{1}{c}{Copula} & \multicolumn{1}{c}{} & \multicolumn{1}{c}{DR} & \multicolumn{1}{c}{H} & \multicolumn{1}{c}{Copula} \\ \hline
		\multirow{6}{*}{$\mathbb{E}(C|x)$} & {Case 1} & 0.25                 &  & -0.03                  & -0.04                   & -0.41                   &                      & 0.05                   & 0.04                    & 0.20                    \\
		&                         & 0.50                 &  & 0.06                   & -0.03                   & -0.43                   &                      & 0.10                   & 0.08                    & 0.27                    \\
		&                         & 0.75                 &  & 0.37                   & 0.09                    & -0.30                   &                      & 0.34                   & 0.18                    & 0.30                    \\
		& {Case 2} & 0.25                 &  & 0.02                   & 0.02                    & -0.10                   &                      & 0.01                   & 0.01                    & 0.02                    \\
		&                         & 0.50                 &  & -0.02                  & -0.02                   & -0.18                   &                      & 0.02                   & 0.02                    & 0.05                    \\
		&                         & 0.75                 &  & 0.02                   & -0.01                   & -0.23                   &                      & 0.03                   & 0.03                    & 0.08                    \\ \hline
		\multirow{6}{*}{$Std(C|x)$}        & {Case 1} & 0.25                 &  & 0.11                   & -0.28                   & -0.14                   &                      & 0.04                   & 0.09                    & 0.04                    \\
		&                         & 0.50                 &  & 0.43                   & -0.35                   & 0.38                    &                      & 0.24                   & 0.14                    & 0.18                    \\
		&                         & 0.75                 &  & 0.36                   & -1.33                   & 0.65                    &                      & 0.25                   & 1.81                    & 0.49                    \\
		& {Case 2} & 0.25                 &  & 0.04                   & 0.00                    & -0.13                   &                      & 0.01                   & 0.01                    & 0.03                    \\
		&                         & 0.50                 &  & 0.01                   & -0.05                   & -0.18                   &                      & 0.02                   & 0.02                    & 0.06                    \\
		&                         & 0.75                 &  & 0.07                   & -0.06                   & -0.16                   &                      & 0.04                   & 0.03                    & 0.06                    \\ \hline
		\multirow{6}{*}{$ES_{0.95}(C|x)$}        & {Case1} & 0.25                     &                      & 0.77                   & -2.58                   & -0.14                   &                      & 1.01                   & 6.80                    & 0.32                    \\
		&                        & 0.5                      &                      & 2.24                   & -3.98                   & 1.89                    &                      & 5.86                   & 16.11                   & 4.32                    \\
		&                        & 0.75                     &                      & 1.75                   & -9.88                   & 2.26                    &                      & 5.18                   & 98.21                   & 6.69                    \\
		& {Case2} & 0.25                     &                      & 0.24                   & 0.02                    & -0.21                   &                      & 0.20                   & 0.08                    & 0.18                    \\
		&                        & 0.5                      &                      & 0.08                   & -0.31                   & -0.27                   &                      & 0.30                   & 0.26                    & 0.36                    \\
		&                        & 0.75                     &                      & 0.34                   & -0.53                   & 0.05                    &                      & 0.72                   & 0.56                    & 0.54                    \\ \hline
		\multirow{6}{*}{$Q_{0.95}(C|x)$}   & {Case 1} & 0.25                 &  & -0.07                  & -0.79                   & -0.46                   &                      & 0.24                   & 0.74                    & 0.39                    \\
		&                         & 0.50                 &  & 0.66                   & -1.34                   & 0.75                    &                      & 0.90                   & 2.02                    & 1.00                    \\
		&                         & 0.75                 &  & 0.08                   & -4.72                   & 1.00                    &                      & 1.09                   & 22.77                   & 1.96                    \\
		& {Case 2} & 0.25                 &  & -0.01                  & 0.25                    & -0.55                   &                      & 0.08                   & 0.12                    & 0.36                    \\
		&                         & 0.50                 &  & -0.12                  & 0.26                    & -0.71                   &                      & 0.18                   & 0.18                    & 0.64                    \\
		&                         & 0.75                 &  & -0.32                  & 0.12                    & -0.98                   &                      & 0.40                   & 0.20                    & 1.22                    \\ \hline
	\end{tabular}
	\begin{minipage}{.8\linewidth} \footnotesize
		\textit{Notes}: Refer to Table \ref{tab: PGB2-C}.
	\end{minipage}
\end{table}

For this DGP, we assume that $F_Z$ and $F_Y$ follow the Poisson and Gamma distributions, respectively, for both parametric models, and the log link function is adopted for both the $g_Z$ and $g_Y$. Therefore, the Gaussian copula model is correctly specified for this simulation exercise, and the hierarchical model is correctly specified in the margin but with a different dependence structure. The comparison results are provided in Table \ref{tab: Copula-C}. The hierarchical model gives the best estimate of the mean value, but the DR Method performs comparatively well. The correctly specified Gaussian copula model performs better than the hierarchical model for the standard deviation, ES, and quantile in case 1, while the hierarchical model has better performance in most scenarios in case 2. The DR approach reveals a great advantage in estimating the quantile, and it is reasonably comparable with the better model for other quantities.

\paragraph{Truncated Bivariate Normal DGP}
\begin{table}[H]
	\centering
	\caption{ Monte Carlo Results under Truncated Bivariate Normal DGP\label{tab: Normal-C}}
	\linespread{1.0}	
	\small	
	\begin{tabular}{llrlrrrrrrr}
		\hline\hline
		&                         & \multicolumn{1}{l}{} &  & \multicolumn{3}{c}{Bias}                                                   & \multicolumn{1}{c}{} & \multicolumn{3}{c}{MSE}                                                    \\ \cline{5-7} \cline{9-11} 
		Quantities                         & Case                    & $x_1$                &  & \multicolumn{1}{c}{DR} & \multicolumn{1}{c}{H} & \multicolumn{1}{c}{Copula} & \multicolumn{1}{c}{} & \multicolumn{1}{c}{DR} & \multicolumn{1}{c}{H} & \multicolumn{1}{c}{Copula} \\ \hline
		\multirow{6}{*}{$\mathbb{E}(C|x)$} & {Case 1} & 0.25                 &  & -0.52                  & 2.60                    & -0.06                   &                      & 0.93                   & 8.28                    & 0.58                    \\
		&                         & 0.50                 &  & -0.66                  & 2.92                    & -0.08                   &                      & 1.25                   & 10.44                   & 0.72                    \\
		&                         & 0.75                 &  & -0.75                  & 3.13                    & -0.19                   &                      & 1.43                   & 11.91                   & 0.74                    \\
		& {Case 2} & 0.25                 &  & -0.11                  & 0.06                    & 0.26                    &                      & 0.07                   & 0.07                    & 0.13                    \\
		&                         & 0.50                 &  & 0.12                   & 0.27                    & 0.46                    &                      & 0.09                   & 0.15                    & 0.28                    \\
		&                         & 0.75                 &  & 0.12                   & 0.29                    & 0.47                    &                      & 0.08                   & 0.16                    & 0.28                    \\ \hline
		\multirow{6}{*}{$Std(C|x)$}        & {Case 1} & 0.25                 &  & -1.39                  & 14.28                   & -1.98                   &                      & 3.18                   & 222.93                  & 4.46                    \\
		&                         & 0.50                 &  & -1.52                  & 15.77                   & -1.98                   &                      & 3.80                   & 271.11                  & 4.62                    \\
		&                         & 0.75                 &  & -1.68                  & 17.39                   & -2.08                   &                      & 4.33                   & 331.55                  & 4.99                    \\
		& {Case 2} & 0.25                 &  & -0.44                  & 1.06                    & -0.41                   &                      & 0.41                   & 1.76                    & 0.29                    \\
		&                         & 0.50                 &  & 0.05                   & 1.54                    & -0.07                   &                      & 0.33                   & 3.00                    & 0.14                    \\
		&                         & 0.75                 &  & 0.21                   & 1.86                    & 0.13                    &                      & 0.31                   & 4.03                    & 0.14                    \\ \hline
		\multirow{6}{*}{$ES_{0.95}(C|x)$}         &  {Case1} & 0.25 &  & 1.49  & 40.83 & -7.16  &  & 19.70 & 1740.44 & 56.91  \\
		&             & 0.5  &  & 1.53  & 44.76 & -7.92  &  & 21.27 & 2100.22 & 68.88  \\
		&             & 0.75 &  & -1.35 & 47.45 & -10.33 &  & 28.76 & 2363.04 & 113.11 \\
		&  {Case2} & 0.25 &  & 0.82  & 4.33  & -0.64  &  & 12.85 & 26.86   & 3.78   \\
		&             & 0.5  &  & 2.21  & 5.03  & -0.33  &  & 24.43 & 34.72   & 4.19   \\
		&             & 0.75 &  & 0.95  & 4.49  & -1.07  &  & 15.79 & 19.61   & 5.01   \\    \hline		
		\multirow{6}{*}{$Q_{0.95}(C|x)$}   & {Case 1} & 0.25                 &  & -2.24                  & 8.05                    & -5.78                   &                      & 18.56                  & 90.73                   & 38.24                   \\
		&                         & 0.50                 &  & -2.45                  & 10.51                   & -5.74                   &                      & 21.73                  & 151.72                  & 38.95                   \\
		&                         & 0.75                 &  & -3.00                  & 12.47                   & -6.40                   &                      & 24.21                  & 205.11                  & 46.76                   \\
		& {Case 2} & 0.25                 &  & -0.52                  & 0.47                    & 0.66                    &                      & 2.11                   & 2.05                    & 1.13                    \\
		&                         & 0.50                 &  & 0.92                   & 1.81                    & 1.63                    &                      & 3.87                   & 5.22                    & 3.53                    \\
		&                         & 0.75                 &  & 0.56                   & 1.51                    & 0.94                    &                      & 3.60                   & 4.08                    & 1.67                    \\ \hline
	\end{tabular}
	\begin{minipage}{.9\linewidth} \footnotesize
		\textit{Notes}: Refer to Table \ref{tab: PGB2-C}.
	\end{minipage}
\end{table}
The distributions of $Z$ and $Y$ in this DGP are not the generally adopted distribution families; it is thus difficult for the parametric models to specify the appropriate distributions directly. Based on the sample information, we specify the Poisson and Gamma distributions as the marginal distributions for both parametric models. As shown in Table \ref{tab: Normal-C}, the hierarchical model fails to estimate all quantities properly in case 1, but it can give desirable results in case 2, especially for the mean.  The copula model provides the best estimation of the mean and standard deviation in case 1, while the DR approach greatly outperforms the two parametric models on estimating the quantile and ES in most scenarios.

Overall, the simulation results reveal an advantage of our semiparametric approach on estimating the higher order moments, quantiles, and ES. The parametric models always estimate the mean properly in that they are mean-based regression models, while the DR approach is reasonably comparable to the correctly specified model. In practice, the exact distributional characteristics are never known exactly, and our approach provides an estimation procedure for flexibly modeling the joint distribution of multiple random variables conditional on some covariates.

In Appendix \ref{sec: appendix-B}, we present additional simulation results. First, the 95th conditional quantile, mean, and standard deviation of $Y$ under each DGP are compared. Besides, another DGP based on the hierarchical model with Negative Binomial and Log Normal distributions are considered. In this section, all comparisons are conducted by looking at three different covariate values. In the appendix, we provide additional comparison results by looking at $1000$ randomly generated covariates from the uniform distributions, which is considered as a cohort. For each DGP, the 95\% VaR and ES of $C$ for this cohort are explored in both cases.

\section{Real Data Analysis}
\label{sec:real}
To test the empirical application of our method, we analyze motor third-party liability policies from an unknown French insurance company.
In non-life insurance, the Collective Risk Model (CRM) has become
one of the most crucial decision-making models. With recent developments, different dependent structures between the
frequency and severities are accommodated to extend the traditional
CRM \citep[see][]{garrido2016generalized, czado2012mixed}. The hierarchical and copula strategies for mixed bivariate
modeling are widely applied by considering the claim frequency as
the discrete outcome and the individual or the average claim severity
as the continuous outcome. As discussed previously, the proposed DR
approach can also be applied to study the insurance data without worrying about the model's specifications.
For the real data analysis, we compare our method with two popular
dependent CRMs: the hierarchical model and the Gaussian copula regression model.

\subsection{Data Description }
The French Motor Third-Part Liability data we used are publicly available
(R-Package \textit{CASdatasets}).
The data comprises 413,169 observations; each consists of a set of characteristics associated with the policyholder and their past claim experience. Specifically,
the datasets consist of the number of claims (frequency), individual claim amounts, and several rating factors (listed in Table \ref{tab:Description-of-covariates}) for each policyholder observed mostly in one year. 
In this application, the discrete and continuous outcomes of interest are the claim frequency $Z$ and the average severity $Y$. The average severity for each policyholder is created by taking an average of the claim amounts over the number of claims made. In addition, we obtain the corresponding aggregate claim amounts $S$ of each policyholder for analysis. 
We create dummies for the non-ordered categorical variables for model construction, e.g., the car brands and administrative regions, and we have a total of $20$ covariates. In our analysis, all rating factors are considered covariates $X$ when modeling the claim frequency and average severity in different estimation models.
\vspace{-0.2cm}
\begin{table}[H]
	\label{tab:x}
	\centering
	\linespread{1.0}
	\small
	\caption{Description of rating factors\label{tab:Description-of-covariates}}
	\begin{tabular}{lll}
		\hline \hline
		Covariate & & \multicolumn{1}{l}{ Description } \\
		\hline Car Power & & The power of the (ordered categorical). \\
		Car Age & & The age of the car, in years. \\
		Car brand & & The brand of the car, grouped into seven categories. \\
		Car gas & & The gas of the car, either Diesel or regular. \\
		Driver age & &The driver age, in years (in France, people can drive a car at 18). \\
		Region & &The region of the policy in France, grouped into five categories). \\
		Density & &The density of inhabitants in the city the policyholder lives in.\\
		\hline
	\end{tabular}
\end{table}
A couple of conclusions can be drawn from a preliminary statistical analysis of the dataset. First, the data is heavily concentrated on zero, with 397,779 (or 96.28\%) policyholders not having made any claims at all. For the policyholders who made claims, the distribution of the average severity is very skewed (refer to panel (a) in Figure \ref{fig:Histograms-for-Y}), that the mean of the distribution exceeds the 75\% quantile and the median  is close to its $75\%$
quantile. This phenomenon is well documented in the literature \cite{YangLu2020NCEf}. In addition, the bimodal shape implies that grasping its behavior fully would be difficult with traditional parametric approaches.  

\subsection{Model Specifications and Comparisons}

For the hierarchical and Gaussian copula models, we use a GLM with Poisson distribution and log link function for modeling the conditional distribution of the claim frequency $Z$. Given that the empirical distribution of the average severity $Y$ is right-skewed and long-tailed, a GLM with GB2 distribution and identical link function is applied for modeling its conditional distribution in both parametric models. In the proposed DR approach, we use the logit link function and identical transformation for both the frequency and the average severity. The support of $Z$ is $\{0,1,\ldots,4\}$ in this dataset, and the discretization points for estimating the distribution of $Y$ are chosen as the $0.1\%, 0.2\%,\ldots,100\%$ quantiles of positive $\{Y_i\}_{i=1}^{n_1}$.

To run the model comparison, we randomly separate the total dataset into the training set with $n_{1}=$ 300,000 policyholders and the validation set with $n_{2}=$ 113,169 policyholders. In the training set, there are only 12,139 policyholders who had made claims. 

\subsubsection{In-Sample and Out-of-Sample Performance}
\begin{figure}[H]
	\captionsetup[subfigure]{aboveskip=-3pt,belowskip=0pt}
	\centering
	\caption{Histograms for the Empirical and Estimated Distribution of positive $Y$\label{fig:Histograms-for-Y}}
	\begin{subfigure}[b]{0.40\textwidth}
		\centering
		\caption{Empirical}     		
		\includegraphics[width=0.95\textwidth]{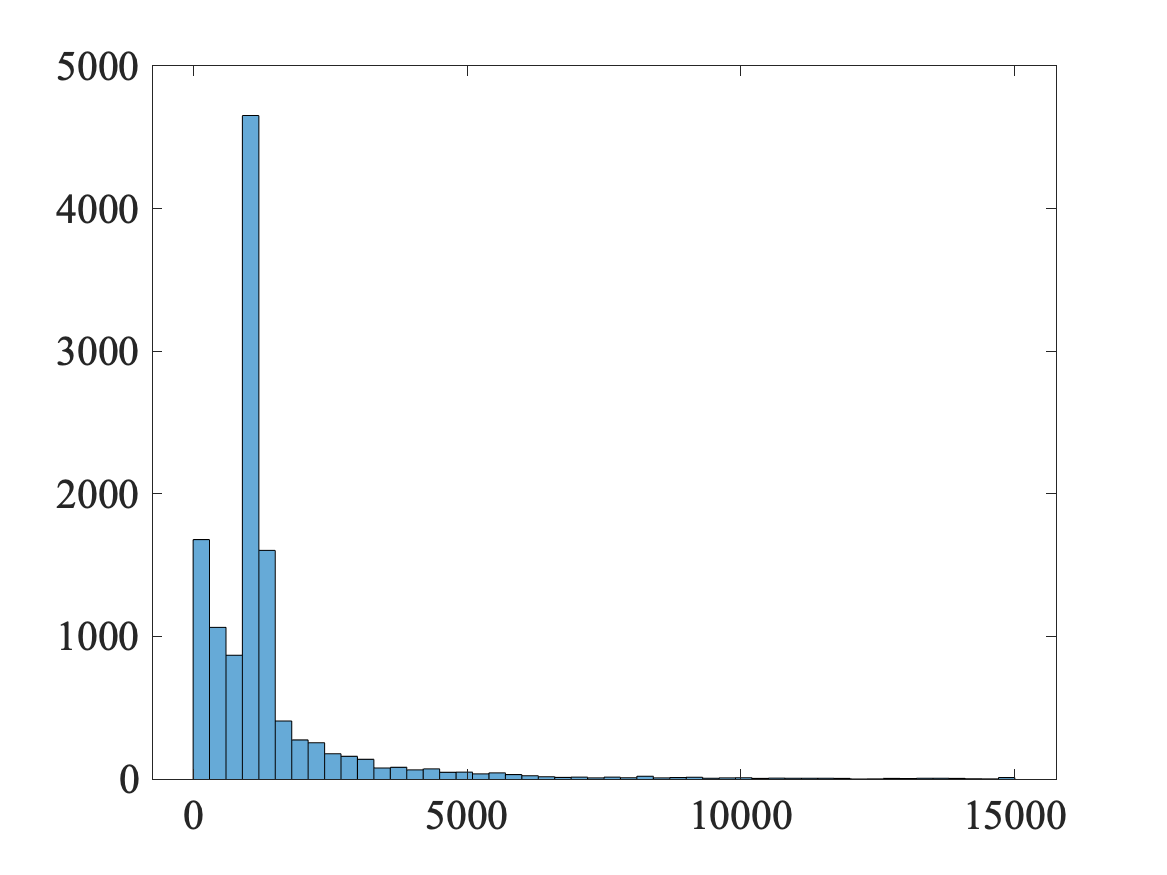}      		
		\label{H-empirical}
	\end{subfigure}
	\begin{subfigure}[b]{0.40\textwidth}
		\centering
		\caption{DR}		
		\includegraphics[width=0.95\textwidth]{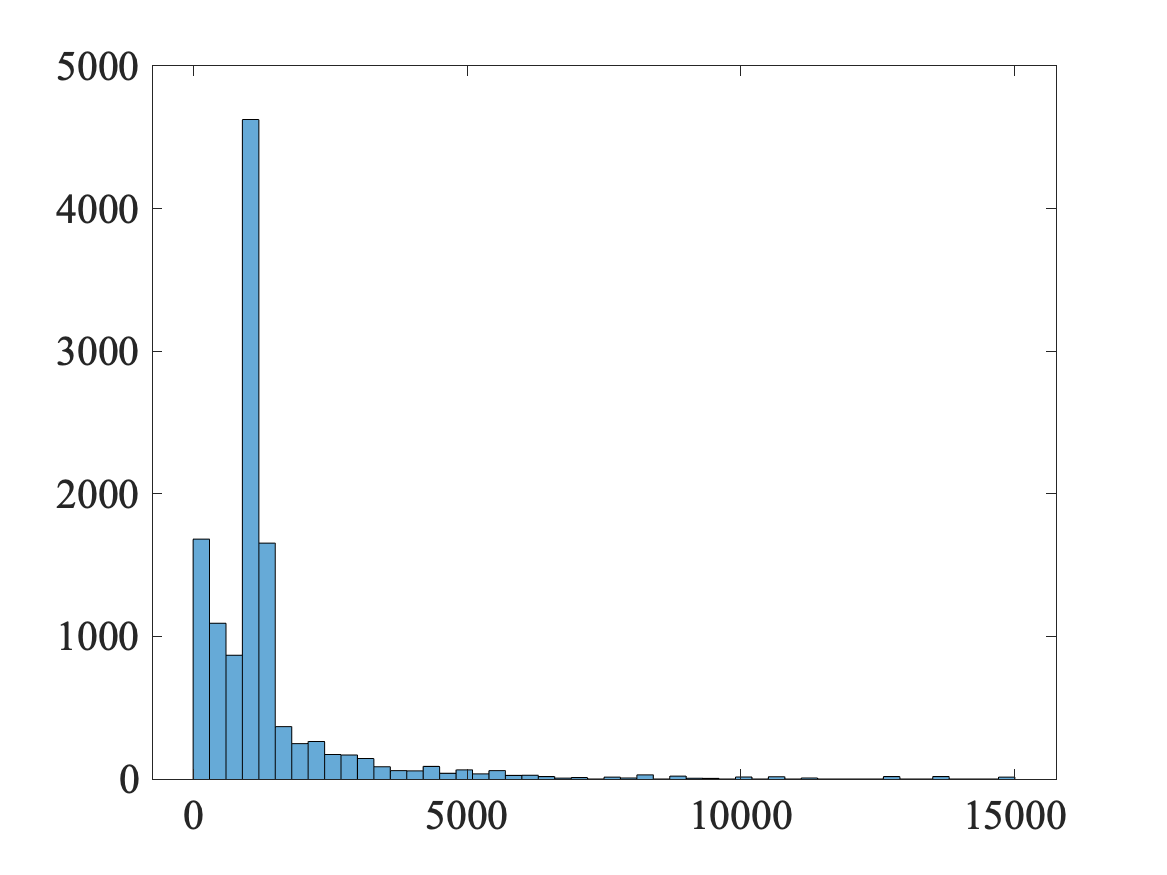}
		\label{H-DR}
	\end{subfigure}
	\hfill
	\begin{subfigure}[b]{0.40\textwidth}
		\centering
		\caption{Hierarchical}		
		\includegraphics[width=0.95\textwidth]{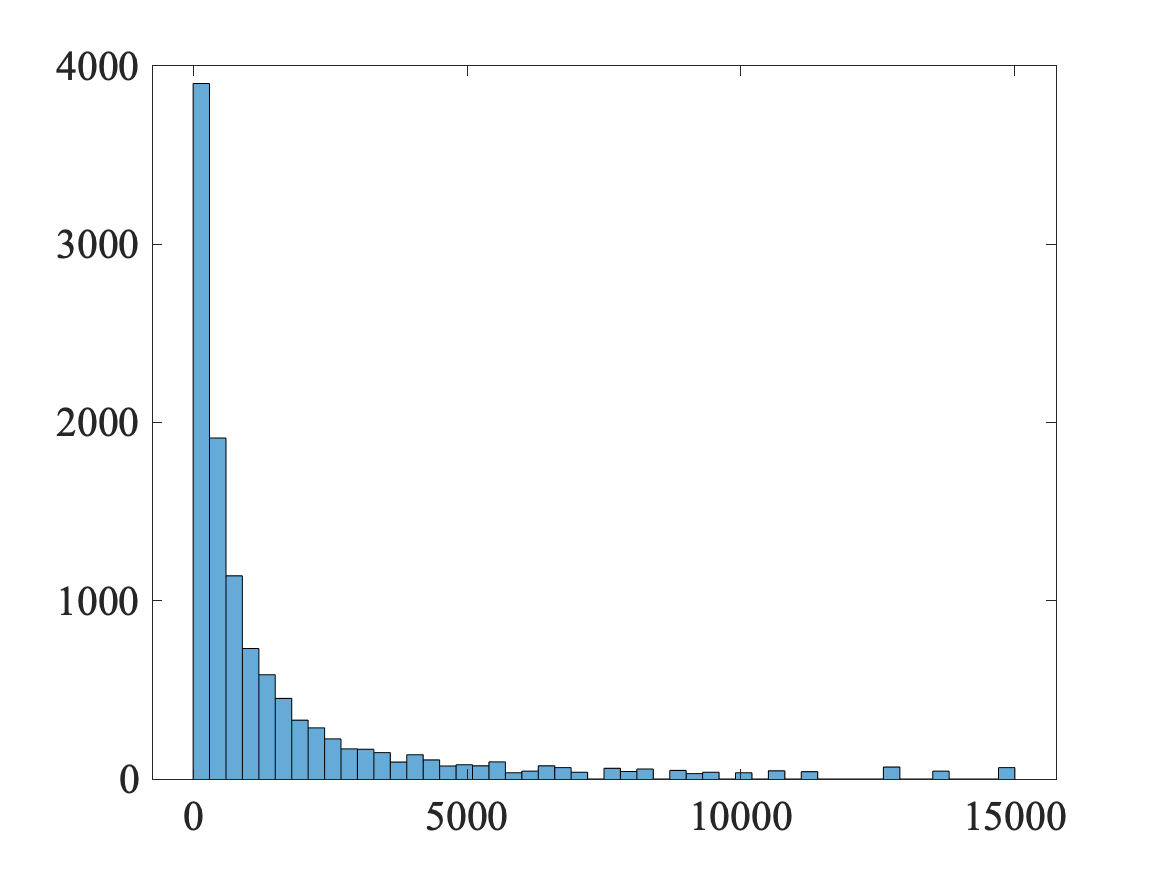}
		\label{H-PG}
	\end{subfigure}          
	\begin{subfigure}[b]{0.40\textwidth}
		\centering
		\caption{Copula}
		\includegraphics[width=0.95\textwidth]{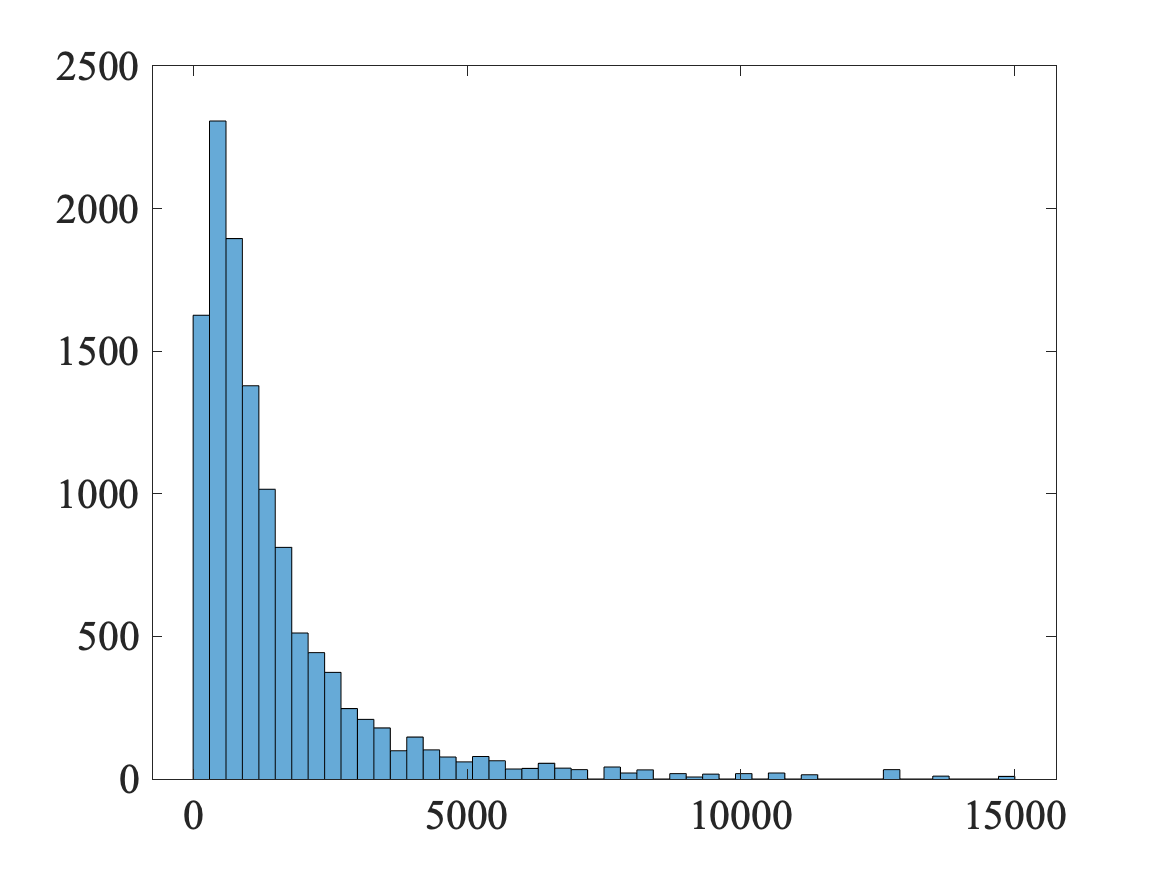}
		\label{H-Copula}
	\end{subfigure}
	\begin{minipage}{.78\linewidth} \linespread{1.0} \footnotesize 
		\textit{Notes}: Panel (a) is constructed for all positive in-sample observations of $Y$, and panels (b)-(d) are constructed based on $n_1$ samples generated from the estimated unconditional distributions $n_1^{-1}\sum_{i=1}^{n_1}\widehat{F}_{Y|X=x_i,Z>0}$ by the DR, P-G and Copula models, respectively. All histograms are constructed with the binsize set as 300.
	\end{minipage}
\end{figure}

We first look at the claim frequency $Z$. Table \ref{tab:Goodness-of-fit-test} reports the observed frequency and the fitted frequency using the
estimated models and the chi-square statistics. The fitted frequency
	is calculated as $n_1\widehat{P}(Z=z)$, where the estimated probabilities $\widehat{P}(Z=z)$ can be obtained from the $\widehat{F}_{Z|X=x_i}$ as follows,
	$$
	\widehat{P}(Z=z)=\frac{1}{n_1}\sum _{i=1}^{n_1}\left(\widehat{F}_{Z|X}(z|x_i)-\widehat{F}_{Z|X}(z-1|x_i)\right).
	$$
The much smaller test statistic of the proposed method suggests a better performance than the global assumption of the Poisson distribution. More importantly, unlike the parametric approach here, the DR approach can capture the upper tail of the distribution that helps insurers oversee and manage their claim-handling expenses. 
\begin{table}[H]
	\centering
	\linespread{1.0}
	\small
	\caption{Goodness-of-fit Test for Claim Frequency\label{tab:Goodness-of-fit-test}}
	\begin{tabular}{crrr}
		\hline \hline
		$z$ & {Empirical} & {DR} & {Poisson} \\
		\hline 0 & 287,861 & 287,861 & 287,563 \\
		1 & 11,591 & 11,591 & 12,168 \\
		2 & 527 & 527 & 265 \\
		3 & 19 & 19 & 4 \\
		4 & 2 & 2 & 0 \\
		Chi-square statistics & & 0 & 428 \\
		\hline
	\end{tabular} \\
	\begin{minipage}{.53\linewidth} \footnotesize
		\textit{Notes}: The fitted frequencies are calculated as $n_1\widehat{P}(Z=z)$ based on the estimated distribution $\widehat{F}_{Z|X=x_i}$.
	\end{minipage}
\end{table}
Note here $Y$ has a mixed distribution. For the average severity $Y$, we set it at zero where the policyholder did not make a claim, that is, $P(Y=0|X)\equiv P(Z=0|X)$. Thus, for all estimation models, we directly model the univariate conditional distributions of $Z$ and $Y|Y>0$ so as to characterize the conditional joint distribution of $(Y,Z)$.
We look at the empirical distribution of positive $Y$ and the estimated distributions, $n_1^{-1}\sum_{i=1}^{n_1}\widehat{F}_{Y|X=x_i,Z>0}$, where
$$\widehat{F}_{Y|X,Z>0}(y|x):=\frac{\int_{\mathcal{Z}\setminus\{0\}}\widehat{F}_{Y|X,Z}(y|x,z)d\widehat{F}_{Z|X}(z|x)}{1-\widehat{F}_{Z|X}(0|x)},
$$
by the DR, the copula, and the hierarchical models. In Figure \ref{fig:Histograms-for-Y}, we plot the histograms created based on samples generated according to the estimated distributions. All of the distributions are extremely long-tailed, with the maximal sample values of the true data, by the DR, hierarchical, and copula models are 210,837, 222,466, 142,466, and 762,466, respectively. So, we truncate the distribution up to 15,000 in all histograms to reduce the visual distraction from the tail. The results show that the DR method fits the true distribution much better,
while the hierarchical and copula models completely fail to capture
the distribution mode.

In addition, we shall investigate the in-sample and out-of-sample performance on estimating the distribution of the aggregative claim amount $S=Z\cdot Y$, which is a quantity of great interest for any insurer.  The conditional distribution of $S$ is provided in the Example above
and its unconditional distribution is obtained by averaging over the covariate.
To demonstrate the efficiency of our estimated distribution, we shall present the estimated out-of-sample CDF via 300 bootstrap samples, randomly selected through permutations with replacements.

\begin{figure}[H]
	\captionsetup[subfigure]{aboveskip=-3pt,belowskip=0pt}
	\centering
	\caption{In-sample and Out-of-sample ``qq-plot" for CDF of positive $S$\label{fig:S_density}}
	\begin{subfigure}[b]{0.40\textwidth}
		\centering
		\caption{In-sample}           			
		\includegraphics[width=0.95\textwidth]{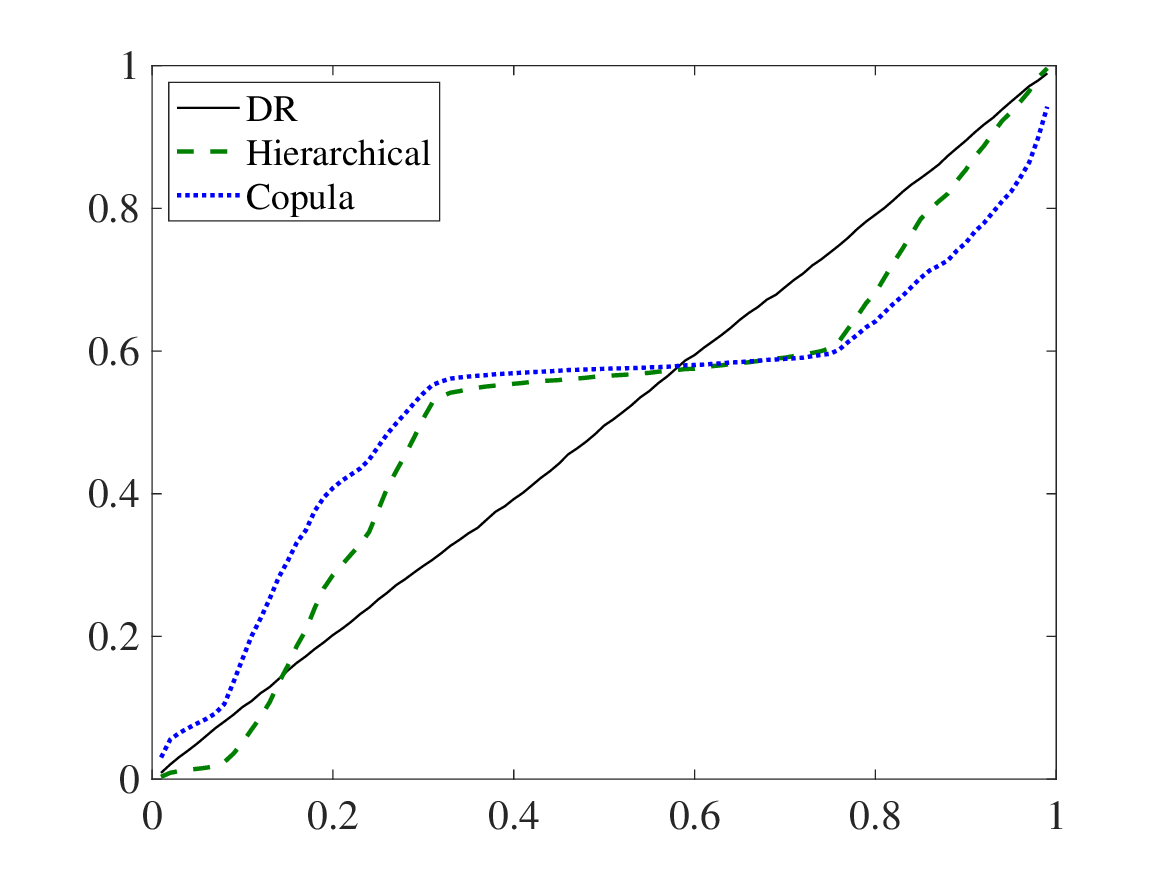}	
		\label{H-empirical}
	\end{subfigure}
	\begin{subfigure}[b]{0.40\textwidth}
		\centering
		\caption{Out-of-sample for DR}		
		\includegraphics[width=0.95\textwidth]{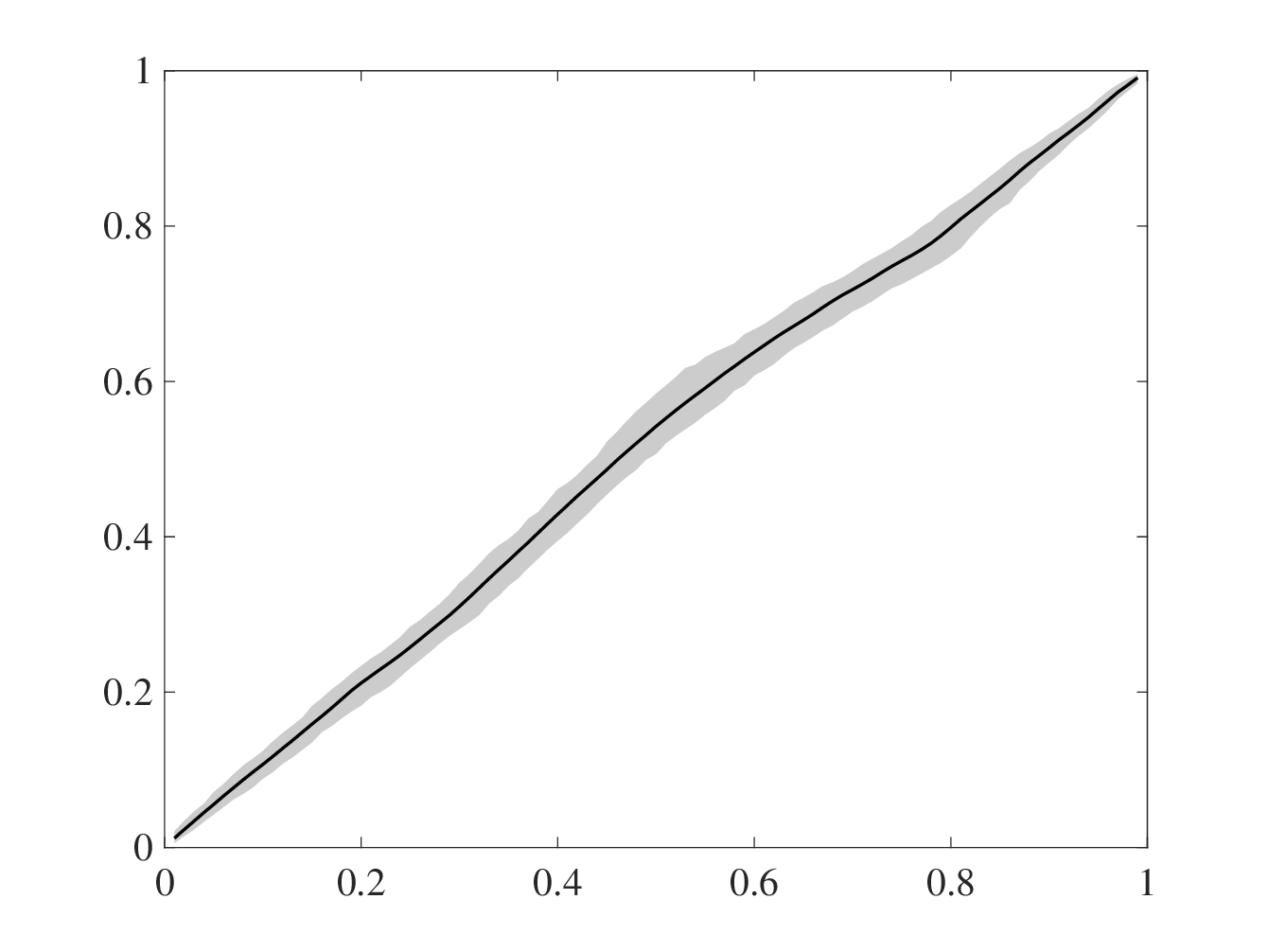}
		\label{H-DR}
	\end{subfigure}
	\hfill	
	\begin{subfigure}[b]{0.40\textwidth}
		\centering
		\caption{Out-of-sample for Hierarchical}		
		\includegraphics[width=0.95\textwidth]{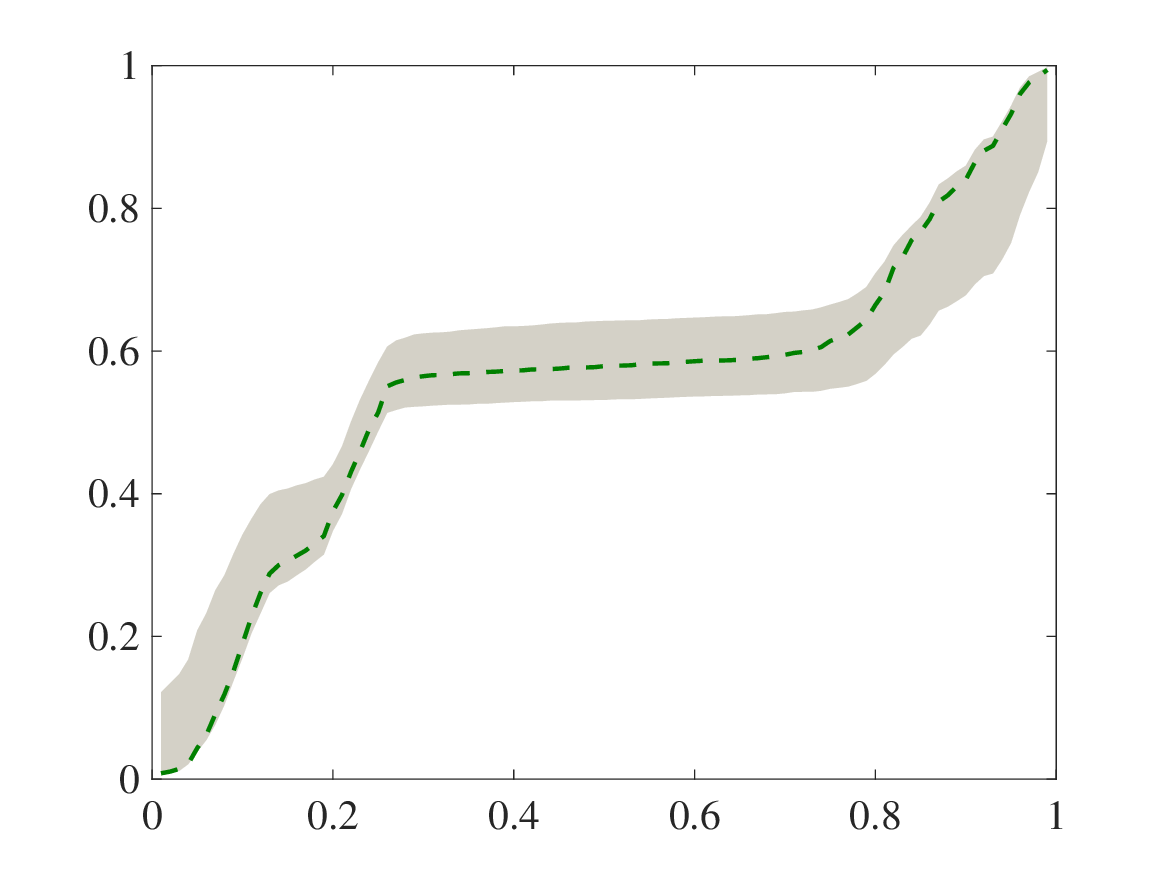}
		\label{H-PG}
	\end{subfigure}          
	\begin{subfigure}[b]{0.40\textwidth}
		\centering
		\caption{Out-of-sample for Copula}		
		\includegraphics[width=0.95\textwidth]{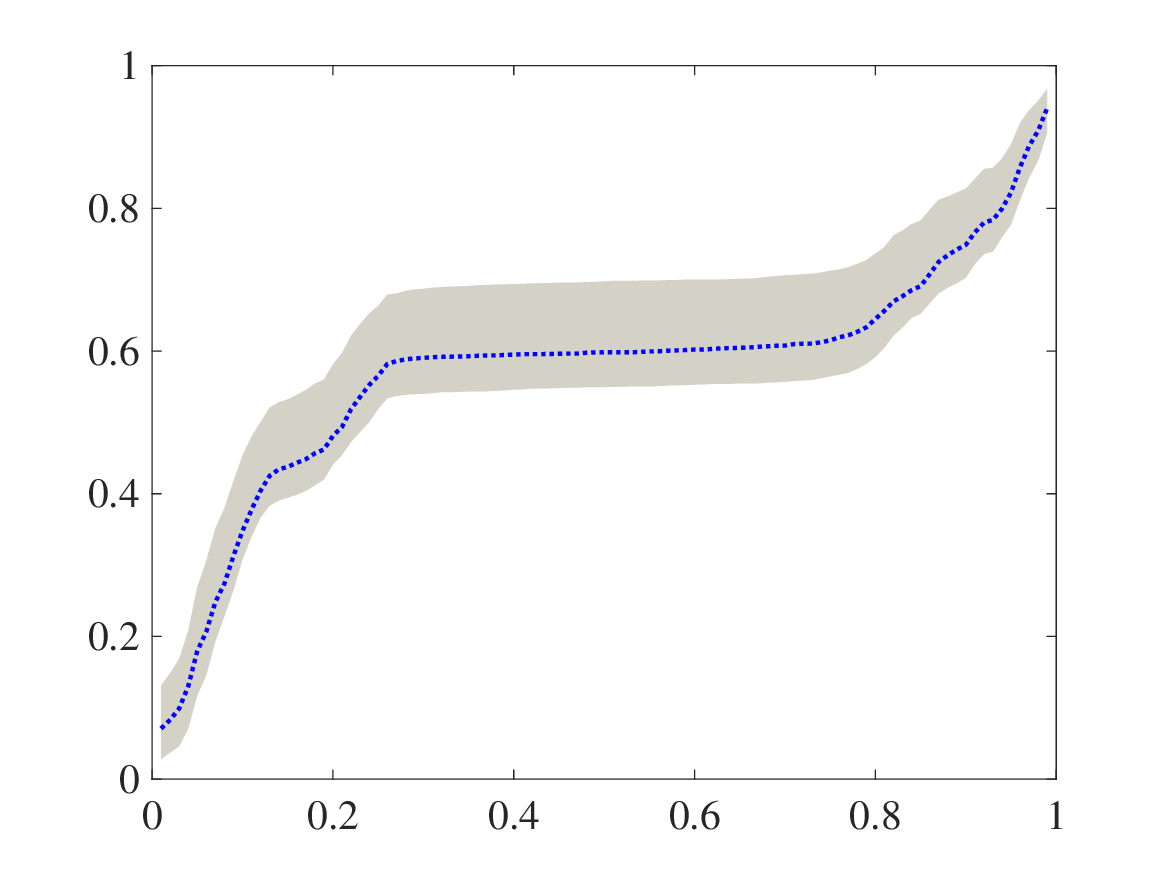}
		\label{H-Copula}
	\end{subfigure}
	\begin{minipage}{.78\linewidth} \linespread{1.0} \footnotesize
		\textit{Notes}: Panel (a) plots the point estimate of the unconditional CDF, $n_1^{-1}\sum_{i=1}^{n_1}\widehat{F}_{S|X=x_i,Z>0}(s|x_i)$, constructed using the three methods against the empirical counterpart (mean of $\1 \{S_{i}\leq s|S_{i}>0\}$) for the in-sample dataset. For the out-of-sample set, panels (b)-(d) present the range (in shadow) and mean (in line) of the estimated CDF's through the three methods based on 300 out-of-sample sets randomly selected using permutations with replacements against their empirical counterparts.
	\end{minipage}
\end{figure}

We plot the estimated unconditional distribution of positive $S$, $n_1^{-1}\sum_{i=1}^{n_1}\widehat{F}_{S|X=x_i,Z>0}(s|x_i)$ over its empirical counterpart, the mean of $\1 \{S_{i}\leq s|S_{i}>0\}$ in Figure \ref{fig:S_density}. 
First, in panel (a), we plot the point estimate of the unconditional CDF of positive $S$ constructed using the three methods against the empirical in-sample CDF. Based on these results, both the hierarchical and the copula models underestimate the CDF function in the lower tail of the distribution and vice versa in the upper tail. 
For insurance companies, these models are often used to forecast, i.e., provide guidelines for pricing and risk management of future policyholders. Hence, out-of-sample performances are important. In panels (b)-(d) of Figure \ref{fig:S_density}, we present the out-of-sample distribution forecast of positive $S$.  The results are quite similar to that of the in-sample ones. The DR approach stands as the preferred approach, with only a slight overestimate of the CDF around the median. 
The results clearly demonstrate the superior performance of our method across the entire distribution and the robustness to the empirical distribution of $S$. 

\subsubsection{Risk Measure Performance}

For risk management purposes, we illustrate the use of our method in the analyses of risk factors on the out-of-sample dataset. The driver's age and the gas type are often critical risk factors in insurance ratemaking and risk management, so in this section, we focus our analysis on four different policyholder cohorts separated by these two factors. 

\begin{table}[H]
	\centering
	\linespread{1.0}
	\small
	\caption{Correlation Coefficients for Different Cohorts\label{tab:Correlation}}
	\begin{tabular}{lrrr}
		\hline\hline
		Correlation     & Pearson & Kendal's tau & Spearman's rho \\ \hline
		Young \& Petrol & -0.025  & -0.077       & -0.014         \\
		Young \& Diesel & -0.043  & -0.042       & -0.040         \\
		Old \& Petrol   & -0.011  & 0.004        & 0.016          \\
		Old \& Diesel   & -0.005  & -0.010       & 0.007          \\ \hline
	\end{tabular}
    \begin{minipage}{.6\linewidth} \linespread{1.0} \footnotesize
    	\textit{Notes}: For the $i$-th policyholder in the out-of-sample dataset with rating factors $x_i$, we generate a pair of sample $(z_i,y_i)$ based on the estimated conditional joint distribution $\widehat{F}_{Z,Y|X}(\cdot,\cdot|x_i)$. The correlation coefficients are calculated based on the generated samples, which are separated into four groups by the driver's age and the gas type.
    \end{minipage}
\end{table}

\begin{figure}[H]
	\captionsetup[subfigure]{aboveskip=-3pt,belowskip=0pt}
	\centering
	\caption{Boxplots of positive $Y$ and $Z$ for Different Policyholder Cohorts\label{fig: scatter plot}}
	\begin{subfigure}[b]{0.40\textwidth}
		\centering
		\caption{Old and Diesel}           			
		\includegraphics[width=0.95\textwidth]{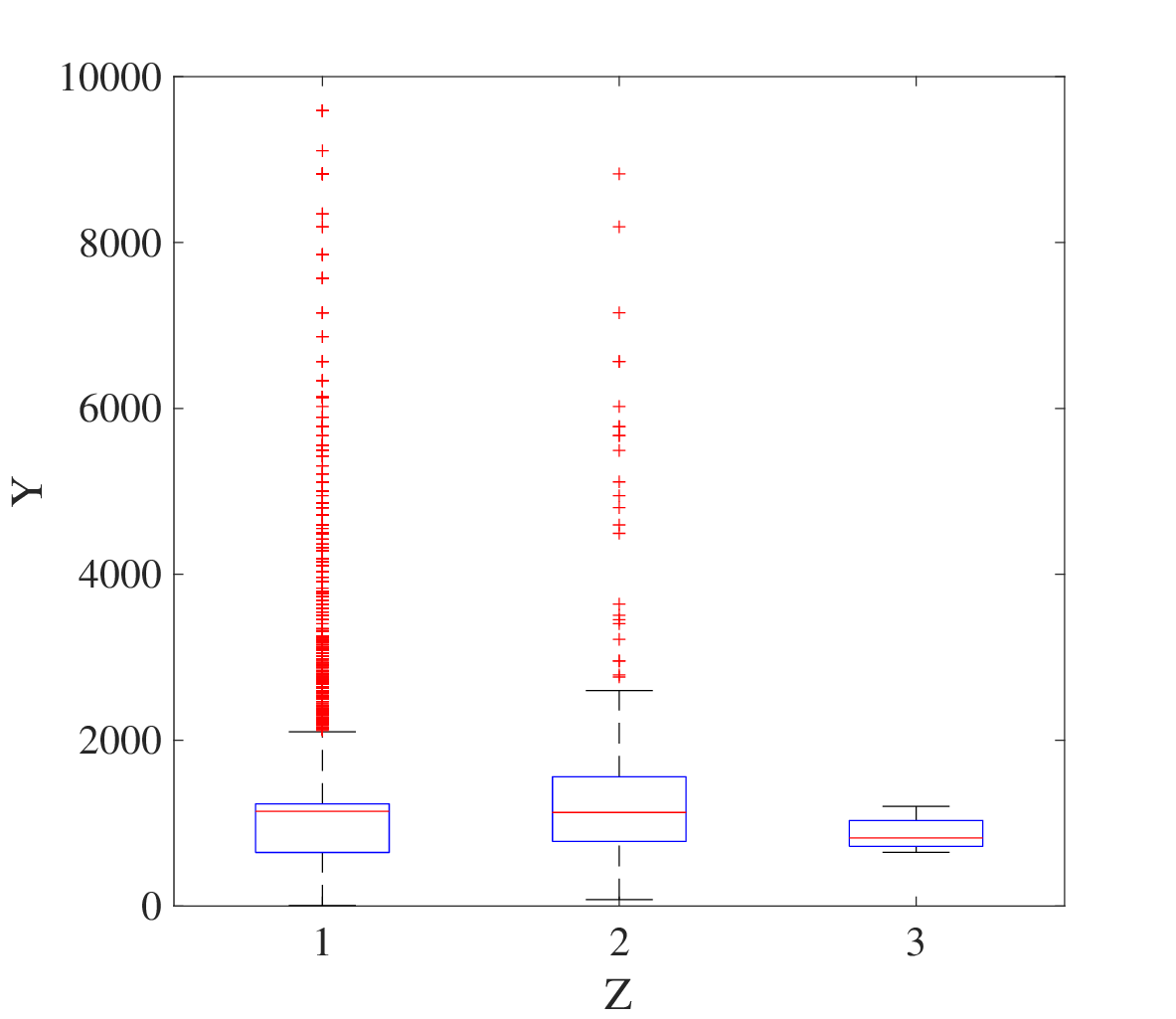}	
		\label{H-empirical}
	\end{subfigure}
	\begin{subfigure}[b]{0.40\textwidth}
		\centering
		\caption{Old and Petrol}		
		\includegraphics[width=0.95\textwidth]{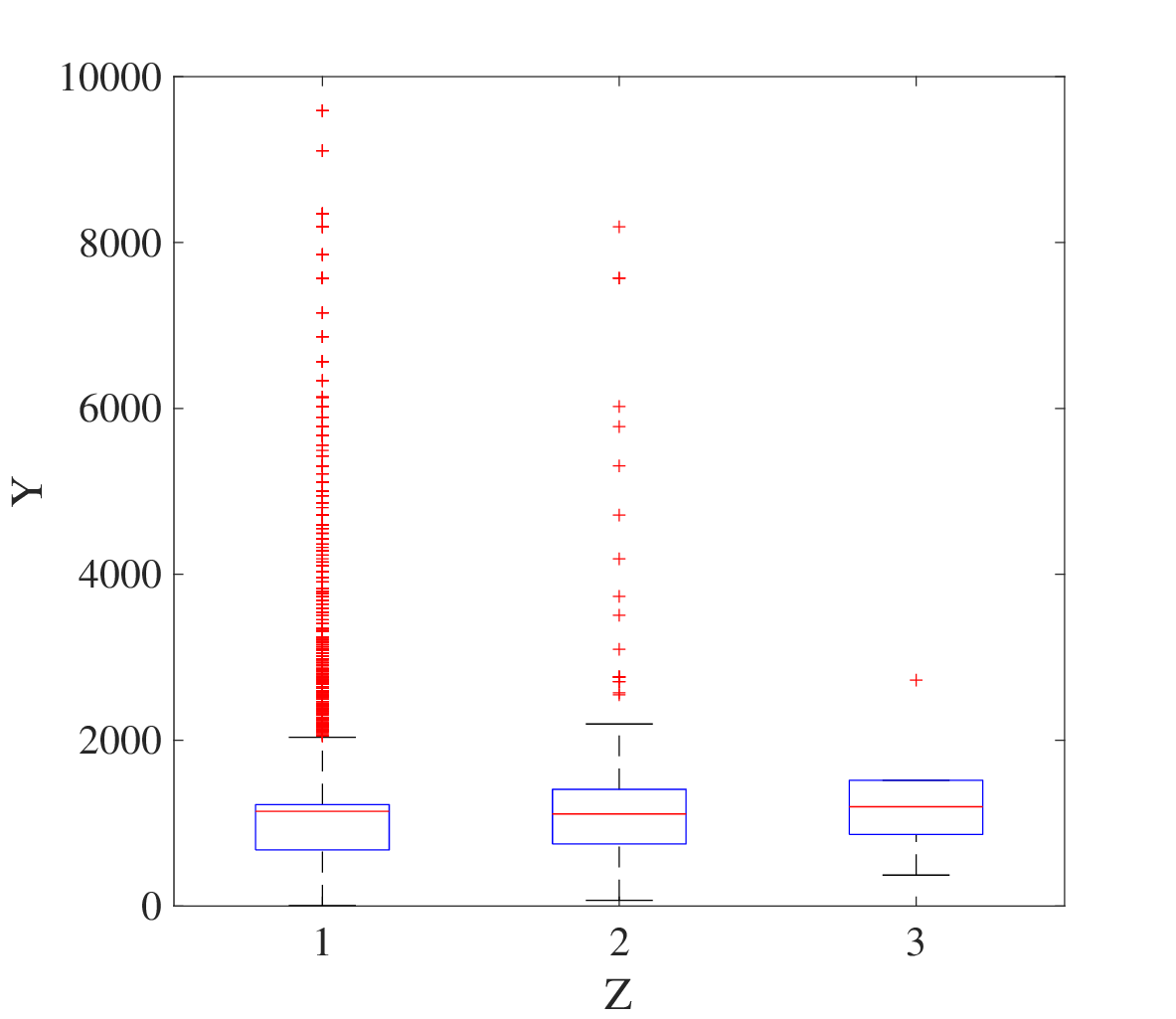}
		\label{H-DR}
	\end{subfigure}
	\hfill	
	\begin{subfigure}[b]{0.40\textwidth}
		\centering
		\caption{Young and Diesel}		
		\includegraphics[width=0.95\textwidth]{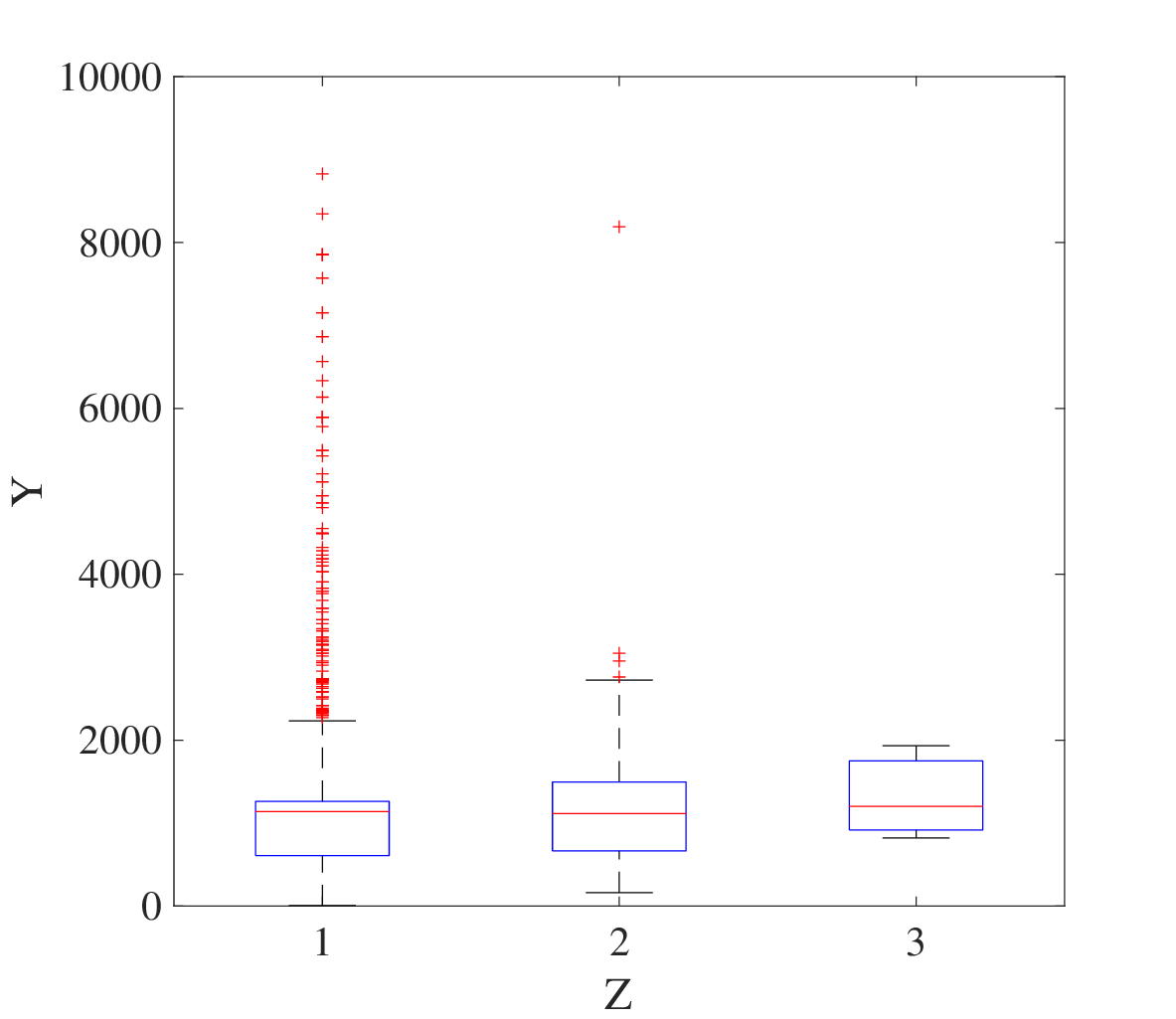}
		\label{H-PG}
	\end{subfigure}          
	\begin{subfigure}[b]{0.40\textwidth}
		\centering
		\caption{Young and Petrol}		
		\includegraphics[width=0.95\textwidth]{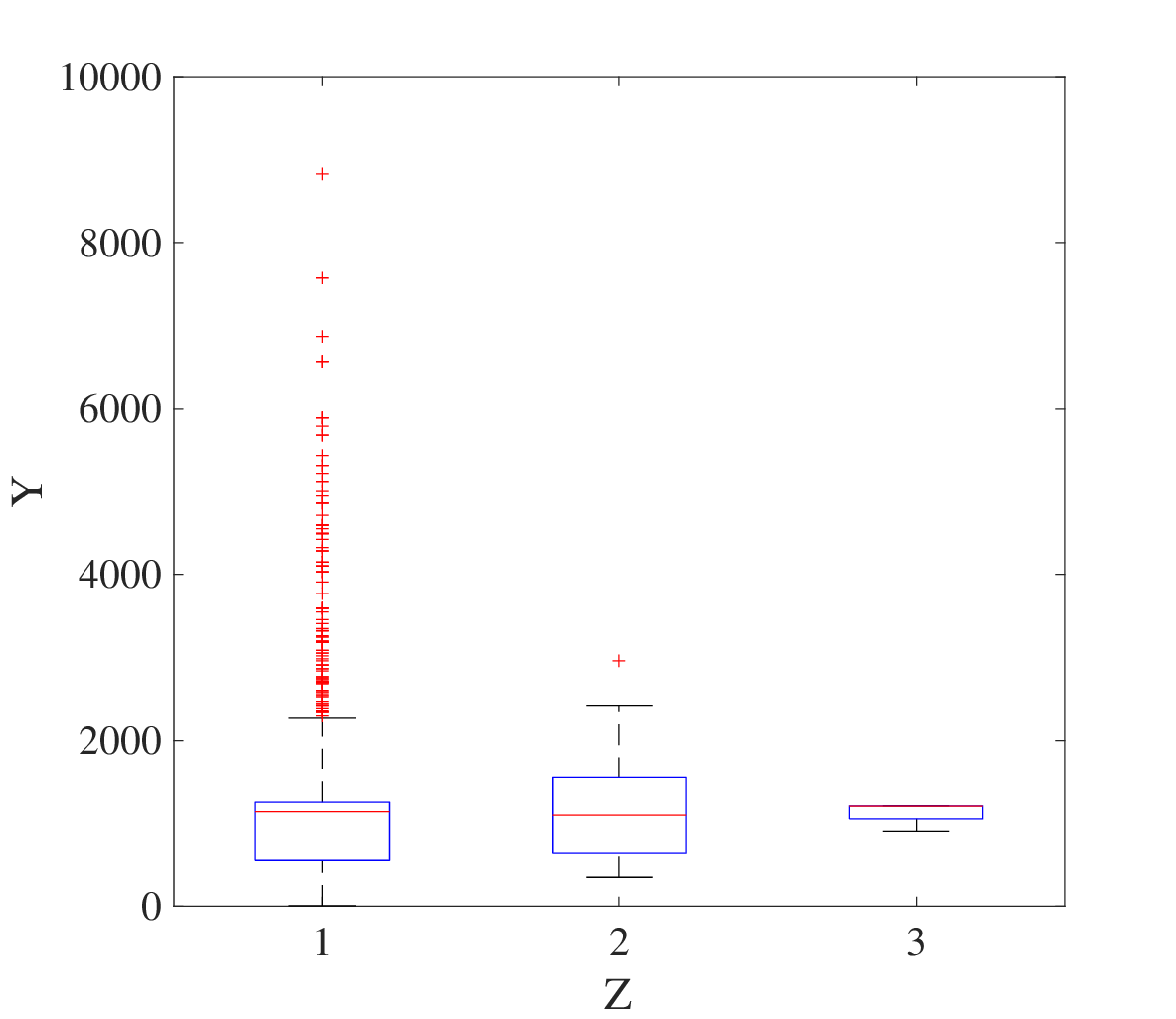}
		\label{H-Copula}
	\end{subfigure}
	\begin{minipage}{.78\linewidth} \linespread{1.0} \footnotesize
		\textit{Notes}: For the $i$-th policyholder in the out-of-sample dataset with rating factors $x_i$, we generate a pair of sample $(z_i,y_i)$ based on the estimated conditional joint distribution $\widehat{F}_{Z,Y|X}(\cdot,\cdot|x_i)$.  The boxplots are constructed based on the generated samples, which are separated into four groups by the driver's age and the gas type.
	\end{minipage}
\end{figure}

First, we explore the dependence between the claim amount and frequency. Based on samples generated via the estimated joint conditional distribution by the DR approach, we compute the Pearson, Kendal's tau, and Spearman's rho correlation coefficients for the four policyholder cohorts, given in Table \ref{tab:Correlation}. All of the correlation coefficients show that there is no significant relationship between these two variables, while this result is misleading in that the probability of a policyholder incurring only one claim is around 0.95 for all cohorts. Furthermore, in Figure \ref{fig: scatter plot}, we visualize the joint distributions for each policyholder cohort using boxplots constructed based on the samples. There is clear evidence that the extreme average severity and frequency are negatively associated, which is consistent with our belief that drivers who incur only one claim are more likely to make extremely large claim amounts, while drivers who file several claims are typically involved in minor accidents. Further comparisons among different cohorts show that drivers over 30 are more likely to be involved in severe accidents than those using petrol cars, while there is no significant difference for drivers using petrol and diesel cars.

We assume that the total cost associated with each policyholder is given as $C=Y\cdot Z+200\cdot Z$, where $200$ is assumed for the fixed claim handling expense when a claim arises. As referred to previously, given the estimated conditional distribution, we can assess the change in the VaR and ES across different cohorts of policyholders based on the quantities $VaR_{\tau}(C|x)$ and $ES_{\tau}(C|x)$. Such analysis allows analysts to identify the cost leaders in the portfolio and make adequate risk management adjustments. 
We demonstrate the out-of-sample risk measures $VaR_{\tau}(C|x)$ and $ES_{\tau}(C|x)$ for $\tau=0.98, 0.99$ computed using our proposed method and that of the competing parametric approaches in Tables \ref{tab:VaR} and \ref{tab: ES}, respectively. In each table, the first two rows corresponding to ``unconditional" are results obtained based on all of the out-of-sample policyholders, and in the remaining rows, we present the results of four different policyholder cohorts.

\begin{table}[H] 
	\centering
	\caption{The Unconditional and Conditional $VaR_{\tau}(C|x)$ \label{tab:VaR}}
	\linespread{1.0}	
	\small
	\begin{tabular}{llccccc}
		\hline\hline
		Cohorts         & $\tau$ & Empirical & DR(EVD) & DR           & Hierarchical & Copula       \\ \hline
		Unconditional   & 0.98   & 1357      & 1342                        & 1340         & 1096         & 787          \\
		&        &           &                             & (1332, 1427) & (1041, 1936) & (679, 1924)  \\
		& 0.99   & 1438      & 1449                        & 1441         & 2030         & 2329         \\
		&        &           &                             & (1346, 1454) & (1137, 2092) & (886, 2769)  \\ \hline
		Young \& Petrol & 0.98   & 1397      & 1346                        & 1330         & 1055         & 491          \\
		&        &           &                             & (1298, 1411) & (920, 1699)  & (382, 767)   \\
		& 0.99   & 1522      & 1493                        & 1496         & 1943         & 1156         \\
		&        &           &                             & (1350, 1614) & (1179, 2145) & (637, 1657)  \\ \hline
		Young \& Diesel & 0.98   & 1383      & 1353                        & 1350         & 1221         & 442          \\
		&        &           & \multicolumn{1}{l}{}        & (1326, 1438) & (1062, 1907) & (352, 652)   \\
		& 0.99   & 1491      & 1621                        & 1642         & 2184         & 901          \\
		&        &           &                             & (1371, 1915) & (1340, 2399) & (558, 1256)  \\ \hline
		Old \& Petrol   & 0.98   & 1106      & 1333                        & 1329         & 1012         & 940          \\
		&        &           & \multicolumn{1}{l}{}        & (1317, 1412) & (930, 1788)  & (777, 2422)  \\
		& 0.99   & 1423      & 1435                        & 1424         & 1915         & 3035         \\
		&        &           &                             & (1339, 1438) & (1074, 2000) & (1091, 3606) \\ \hline
		Old \& Diesel   & 0.98   & 1368      & 1345                        & 1348         & 1160         & 835          \\
		&        &           &                             & (1337, 1431) & (1092, 1998) & (699, 1882)  \\
		& 0.99   & 1441      & 1451                        & 1456         & 2126         & 2319         \\
		&        &           & \multicolumn{1}{l}{}        & (1354, 1484) & (1214, 2219) & (954, 2744)  \\ \hline
	\end{tabular}
	\begin{minipage}{.91\linewidth} \footnotesize
		\textit{Notes}: We study the $VaR_\tau(C|x)$ based on the whole sample (unconditional) and different cohorts (conditional) separated by risk factors ``gas type" and ``age" (if below 30). For each scenario, the empirical values are quantiles of all out-of-sample observations of $C$. For each estimation model, the individual numbers are point estimates computed based on the estimated distribution of $C$, the bracketed numbers are the $95\%$ confidence intervals constructed based on 300 bootstrap results.
	\end{minipage}
\end{table}

We present the point estimates and 95\% confidence intervals of VaR and ES constructed through 300 bootstrap samples for all scenarios in both parametric models. As the support of the claim severity distribution is typically assumed to be infinite in insurance, the proposed approach extrapolates the extreme tail of the conditional distribution by fitting a generalized extreme value distribution, as discussed in Section \ref{sec: Model and Estimation}. The extreme value theory offers a channel for extrapolation outside the range of the available data, as demonstrated in the column labeled "DR(EVD)." However, using bootstrap inference towards the boundary of the support raises challenging theoretical problems. Therefore, in this case, only the point estimates for each scenario are provided, and inference is left for future study. On the other hand, the DR approach proposed on compact support captures the tail behavior of the distribution comparable to the DR with extreme value extrapolation. Additionally, the compact support argument provides the convenience of standard statistical inference via bootstrap samples.

\begin{table}[H] 
	\centering
	\caption{The Unconditional and Conditional $ES_{\tau}(C|x)$ \label{tab: ES}}
	\linespread{1.0}	
	\small
	\begin{tabular}{llccccc}
		\hline\hline
		Cohorts         & $\tau$ & Empirical & DR(EVD) & DR           & Hierarchical & Copula       \\ \hline
		Unconditional   & 0.98   & 2840                          & 2880    & 3031                   & 3081                             & 12083                      \\
		&        &                               &         & (2700, 3402)           & (2817, 3207)                     & (8072, 17937)              \\
		& 0.99   & 4282                          & 4384    & 4676                   & 4670                             & 22818                      \\
		&        &                               &         & (4093, 5347)           & (4187, 4899)                     & (14931, 34370)             \\ \hline
		Young \& Petrol & 0.98   & 3959                          & 3939    & 3286                   & 2914                             & 5774                       \\
		&        &                               &         & (2346, 4539)           & (2345, 3461)                     & (2060, 16300)              \\
		& 0.99   & 6387                          & 5517    & 5159                   & 4380                             & 10728                      \\
		&        &                               &         & (3230, 7550)           & (3378, 5392)                     & (3517, 31476)              \\ \hline
		Young \& Diesel & 0.98   & 3324                          & 3332    & 3354                   & 3261                             & 3887                       \\
		&        &                               &         & (2560, 4771)           & (2664, 3808)                     & (1423, 9765)               \\
		& 0.99   & 5224                          & 5254    & 5284                   & 4901                             & 7138                       \\
		&        &                               &         & (3606, 8044)           & (3827, 5939)                     & (2321, 18916)              \\ \hline
		Old \& Petrol   & 0.98   & 2710                          & 2709    & 2879                   & 2925                             & 15708                      \\
		&        &                               &         & (2445, 3469)           & (2600, 3127)                     & (10315, 23339)             \\
		& 0.99   & 4069                          & 4051    & 4392                   & 4458                             & 29769                      \\
		&        &                               &         & (3474, 5458)           & (3865, 4837)                     & (19179, 44845)             \\ \hline
		Old \& Diesel   & 0.98   & 2702                          & 3126    & 3082                   & 3216                             & 10934                      \\
		&        &                               &         & (2673, 3608)           & (2903, 3440)                     & (6959, 17088)              \\
		& 0.99   & 4025                          & 4842    & 4767                   & 4864                             & 20466                      \\
		&        &                               &         & (3941, 5778)           & (4297, 5261)                     & (12612, 32508)             \\ \hline
	\end{tabular}
	\begin{minipage}{.95\linewidth} \footnotesize
		\textit{Notes}: The table presents point estimates of $ES_{\tau}(C|x)$ for different cohorts. 
	\end{minipage}
\end{table}

First, for $VaR_{\tau}(C|x)$, the results reveal that the proposed approach provides much more accurate point estimates than the hierarchical and copula models. We reiterate that the point estimates obtained by extrapolating the conditional distribution using the extreme value distribution are consistent with the results obtained by using the DR approach alone. Additionally, the DR approach without extrapolation provides narrow confidence intervals that accurately capture the true out-of-sample results in all scenarios. Both parametric models, especially the copula model, tend to underestimate the 98\% VaR and overestimate the 99\% VaR for all cohorts. Empirically, we observe the younger cohort's higher risk profile, and the use of diesel intensifies it. The estimated VaR across the three approaches confirms this increase in risk profile. For $ES_{\tau}(C|x)$, our findings from Table \ref{tab: ES} show that in all cases, the proposed approach provides more accurate point estimates that are closer to the empirical results than the hierarchical model, while the copula model completely fails to estimate the ES properly. The ES is a risk measure that is more sensitive to extreme values compared to VaR. Therefore, the DR approach augmented with the extreme value distribution provides more precise point estimates compared to the DR approach without extrapolation. Furthermore, our results confirm that the risk profile is higher for younger cohorts and the use of diesel intensifies the risk.

\section{Conclusion}
\label{sec: conclusion}
This paper proposes a semiparametric method  based on the
distribution regression approach. The advantage of using this proposed method is three-fold. First, we avoid imposing too rigid parametric assumptions, which makes our approach robust for analyzing real data. Secondly, the covariates are incorporated to influence the whole distribution instead of only affecting the distributions' location parameters. Finally, by including the discrete
outcome as a covariate in the conditional distribution of the continuous outcome, our model captures intricate dependence structures between the two outcomes. While our analysis in this paper focuses on bivariate modeling of one discrete and one continuous outcome, the method can be easily extended to allow any random variables, including mixed distributions. The simulation examples under different scenarios demonstrate the
robustness of our method. The empirical study shows that our method can extract
interesting features of a motor insurance portfolio critical to 
pricing and risk management in real data applications.



\newpage
\setstretch{1}
\bibliographystyle{elsarticle-harv} 
\bibliography{DR_ref}

\newpage
\newpage
	\begin{appendices}
	
	\setcounter{section}{0} 
	\setcounter{figure}{0} 
	\setcounter{table}{0} 
	\setcounter{equation}{0} 
	\setcounter{lemma}{0}\setcounter{page}{1}
	\setcounter{proposition}{0} %
	\renewcommand{\thepage}{A-\arabic{page}}
	\renewcommand{\theequation}{A.\arabic{equation}}
	\renewcommand{\thelemma}{A.\arabic{lemma}} 
	\renewcommand{\theproposition}{A.\arabic{proposition}} 
	\renewcommand\thesection{\Alph{section}} 
	\renewcommand\thesubsection{A.{subsection}}
	\renewcommand\thefigure{\thesection.\arabic{figure}}
	\renewcommand\thetable{\thesection.\arabic{table}}
	
	\setstretch{1.4}
	
	\section{Theoretical Results}\label{sec: appendix-A}

	We define,
	for
	$\theta=(\alpha', \beta')' \in \Theta$,
	\begin{eqnarray*}
		\Psi_{n}(\theta, y, z)
		:= 
		\frac{1}{\sqrt{n}}
		\sum_{i=1}^{n}
		\varphi_{\theta, y, z}(X_{i}, Y_{i}, Z_{i}).
	\end{eqnarray*}
	Notice that 
	$\Psi_{n}(\theta, y, z) =
	n^{-1/2}
	\sum_{i=1}^{n}
	\big[
	\nabla \ell_{i, y}
	\big (
	\alpha
	\big)',
	\nabla \ell_{i, z}
	\big (
	\beta
	\big)'\big]'$
	by definition.

	\vspace{0.5cm}
	\begin{lemma}
		\label{lemma:donsker}
		Suppose that Assumptions A1 and A2 hold. Then,  
		the function class 
		$\big\{\Psi_{n}(\theta, y, z): (\theta, y, z)
		\in \Theta {\times} \mathcal{Y} {\times} \mathcal{Z}\big\}$ is Donsker
		with a square-integrable envelope.

	\end{lemma}
	\begin{proof}
		We define
		the function classes
		\begin{eqnarray*}
			\mathcal{F}_{1}:=\big\{ 
			(x,z) \mapsto P_1(x,z)'\alpha: \alpha \in \mathcal{A} \big\}
			\ \ \mathrm{and} \ \ 
			\mathcal{F}_{2}:=\big\{x \mapsto P_2(x)'\beta: \beta \in \mathcal{B}\big\},    
		\end{eqnarray*}
		and also 
		$\mathcal{G}_{1}:=\big\{y \mapsto \1\{y \le v\}: v \in \mathcal{Y}\big\}$
		and
		$\mathcal{G}_{2}:=\big\{z \mapsto \1\{z \le w\}: w \in \mathcal{Z}\big\}$.
		Lemma 2.6.15 of
		\cite{van1996weak}
		shows that 
		$\mathcal{F}_{1}$,
		$\mathcal{F}_{2}$,
		$\mathcal{G}_{1}$
		and 
		$\mathcal{G}_{2}$
		are VC-subgraph classes.
		Given the transformations $P_1: \mathcal{X}{\times}\mathcal{Z}\rightarrow\mathbb{R}^{d_1}$, and $P_2: \mathcal{Z}\rightarrow\mathbb{R}^{d_2}$, we write
		$P_1(x,z)\equiv[P_1^{(1)}(x,z), \dots, P_1^{(d_{1})}(x,z)]'$ and $P_2(z)\equiv[P_2^{(1)}(z), \dots, P_2^{(d_{2})}(z)]'$. Then,
		the function classes  
		$\{P_1^{(r)}(x,z): r=1, \dots, d_1\}$
		and
		$\{P_2^{(l)}(z): l=1, \dots, d_2\}$
		are also 
		VC-subgraph classes.
		Let
		$\mathcal{H}:=\mathcal{H}_{1}\cup\mathcal{H}_{2}$,
		where
		\[
		\mathcal{H}_{1}:=
		\big\{[\mathcal{G}_{1}-\Lambda(\mathcal{F}_{1}) ]R(\mathcal{F}_{1})P_1^{(r)}(X,Z)
		: r = 1, \dots, d_{1}
		\big\}\]
		and
		\[\mathcal{H}_{2}:=
		\big\{[\mathcal{G}_{2}-\Lambda(\mathcal{F}_{2}) ]R(\mathcal{F}_{2})P_2^{(l)}(Z)
		: l = 1, \dots, d_{2}
		\big\},\]
		with 
		$\Lambda(\cdot)$
		being the link function
		and
		$R(\cdot)=\lambda(\cdot)/\{\Lambda(\cdot)[1-\Lambda(\cdot)]\}$.
		The class $\mathcal{H}$ consists of a Lipschitz transformation
		of 
		VC-subgraph classes
		with Lipschitz coefficients 
		and 
		bounded from above by $\| P_1(X,Z) \|$ or $\|P_2(Z)\|$
		up to some constant factor.
		Also, an envelop function for $\mathcal{H}$
		is 
		bounded from above by $\| P_1(X,Z) \|$ or $\|P_2(Z)\|$
		up to some constant factor
		and thus is square integrable under Assumption A1.
		A Lipschitz composition of a Donsker class is a Donsker class
		Van der Varrt \cite[][19.20]{van2000asymptotic}.         
	\end{proof}
	\vspace{0.5cm}

	In the following lemma, we will consider only the estimator depending on $y \in \mathcal{Y}$.
	The same or simpler argument can prove the same result
	for the estimator based on $z \in \mathcal{Z}$
	because $\mathcal{Z}$ consists of finite points.
	For simplicity, let
	$\theta_{0}(y):= \alpha_{0}(y)$
	and
	$\hat{\theta}(y)$
	be the the estimator 
	of $\theta_{0}(y)$.
	The corresponding minimization problem is
	based on 
	the log-likelihood
	$n^{-1}\sum_{i=1}^{n}\ell_{i, y}(\theta)$
	and 
	$\ell_{y}(\theta)$,
	those of which satisfy Assumptions A1-A5.
	For notational simplicity,   
	we define a localized objective function,
	\begin{eqnarray*}
		Q_{n, y}(\delta)
		:=
		\frac{1}{n}\sum_{i=1}^{n}
		\big \{ 
		\ell_{i, y}
		\big (\theta_{0}(y) + n^{-1/2}\delta\big)
		-
		\ell_{i, y}
		\big (\theta_{0}(y)\big)
		\big \}.
	\end{eqnarray*}
	Then, we can write the estimator 
	$\hat{\delta}(y) := \sqrt{n}
	\big (
	\hat{\theta}(y) - \theta_{0}(y)
	\big)$
	as the solution for 
	$  \max_{\delta \in \R^{d_1}}
	Q_{n, y}
	(\delta)
	$.

	\vspace{0.5cm}
	\begin{lemma}
		\label{lemma:argmin}
		Suppose that Assumptions A1-A5 hold. Then,
		we have, uniformly in $y \in \mathcal{Y}$,
		\begin{eqnarray*}
			\sqrt{n}
			\big (
			\hat{\theta}(y)
			-
			\theta_{0}(y)
			\big )
			=
			-
			H_{0, y}^{-1}
			\Psi_{n}
			\big (\theta_{0}(y), y\big) 
			+ o_p(1),
		\end{eqnarray*}
		where 
		$\Psi_{n}(\theta_{0}(y), y):=
		n^{-1/2}
		\sum_{i=1}^{n}
		\nabla \ell_{i, y}
		\big (
		\theta_{0}(y)  
		\big)$.
		
	\end{lemma}
	\begin{proof}

		Let
		$M$ be a finite positive constant.
		Because
		the map 
		$\delta \mapsto Q_{n, y}(\delta)$
		is twice continuously differentiable
		under Assumption A2,
		we can show that 
		$
		n
		Q_{n,y}(\delta)
		=
		\delta'
		\Psi_{n}(\theta_{0}(y), y)             
		+
		\delta'
		n^{-1}
		\sum_{i=1}^{n}
		\nabla^{2}
		\ell_{i, y}
		\big (\theta_{0}(y) \big)
		\delta
		/2
		+
		o(n^{-1}\|\delta\|^2)
		$
		uniformly
		in $y \in \mathcal{Y}$,
		for each fixed $\delta$
		with $\|\delta \| \le M$.
		Also,
		we can show that 
		$
		n^{-1}
		\sum_{i=1}^{n}
		\nabla^{2}
		\ell_{i,y}
		\big (\theta_{0}(y) \big)
		\to^p
		H_{0, y}
		$
		uniformly in $y \in \mathcal{Y}$,
		by the uniform law of large numbers.
		Thus, for each $\delta$
		with $\|\delta \| \le M$,
		we can show that 
		$
		\sup_{y \in \mathcal{Y}}
		\big |
		Q_{n,y}(\delta)
		-
		\widetilde{Q}_{n,y}(\delta)
		\big |
		=
		o_p(n^{-1})
		$,
		where
		\begin{eqnarray*}
			n
			\widetilde{Q}_{n,y}(\delta)
			=
			\delta'
			\Psi_{n}
			\big(\theta_{0}(y), y \big)
			+
			\frac{1}{2}
			\delta'
			H_{0, y}
			\delta.
		\end{eqnarray*} 
		The convexity lemma
		\citep[see][]{pollard1991asymptotics, kato2009asymptotics}
		extends
		the point-wise convergence 
		with respect $\delta$
		to the uniform converges and thus,
		under Assumption A2,
		\begin{eqnarray}
			\label{eq:Q2}
			\sup_{y \in \mathcal{Y}}
			\sup_{\delta : \|\delta \| \le M}
			\big |
			Q_{n,y}(\delta)
			-
			\widetilde{Q}_{n,y}(\delta)
			\big |
			=
			o_p(n^{-1}).
		\end{eqnarray}

		Let 
		$\tilde{\delta}(y)
		:=
		-
		H_{0, y}^{-1}
		\Psi_{n}
		\big (\theta_{0}(y), y\big) 
		$,
		which maximizes
		$\widetilde{Q}_{n,y}(\delta)$.
		Then, 
		simple algebra can show that,
		for any $\delta$ and for some constant $c>0$,
		\begin{eqnarray}
			\label{eq:lbd1}
			\widetilde{Q}_{n,y}\big(\tilde{\delta}(y) \big)
			-
			\widetilde{Q}_{n,y}\big (\delta \big)
			=
			-
			\frac{1}{2 n}
			\big (
			\tilde{\delta}(y) - \delta  
			\big )'
			H_{0, y}
			\big (
			\tilde{\delta}(y) - \delta  
			\big )
			\ge 
			\frac{c}{2 n}
			\| 
			\tilde{\delta}(y) - \delta  
			\|^2,
		\end{eqnarray}
		where
		the last inequality is
		due to that
		$H_{0, y}$
		is negative definite
		under Assumption A4.  
		For any subset $D$ including $\tilde{\delta}(y)$,
		an application of the triangle inequality obtains 
		\begin{eqnarray}
			2
			\sup_{\delta \in D}
			|
			Q_{n, y}
			(\delta)
			-
			\widetilde{Q}_{n, y}
			(\delta)
			|
			&\ge& \notag
			\sup_{\delta \in D}
			\big \{
			\widetilde{Q}_{n,y}\big(\tilde{\delta}(y) \big)
			-
			\widetilde{Q}_{n,y}\big (\delta \big)
			\big \}\\
			&& - \label{eq:tri5}
			\sup_{\delta \in D}
			\big \{
			Q_{n, y}
			\big (\tilde{\delta}(y)\big)
			-
			Q_{n, y}
			(\delta)
			\big\}.
		\end{eqnarray}
		
		Let $\eta>0$ be an arbitrary constant.
		Because of the concavity under Assumption A2,
		difference quotients
		satisfy that,
		for any $\lambda > \eta$
		and
		for any $v \in S^{d_1}$
		with the unit sphere  
		$S^{d_1}$
		in $\R^{d_1}$,
		\begin{eqnarray*}
			\frac{
				Q_{n, y}
				\big (
				\tilde{\delta}(y)
				+
				\eta v
				\big)
				-
				Q_{n, y}
				\big(
				\tilde{\delta}(y)
				\big)
			}{
				\eta
			}
			\ge
			\frac{
				Q_{n, y}
				\big(
				\tilde{\delta}(y)
				+
				\lambda v
				\big)
				-
				Q_{n, y}
				\big(
				\tilde{\delta}(y)
				\big)
			}{
				\lambda
			}.
		\end{eqnarray*}
		This inequality with
		a set $D_{n, \eta}(y):=
		\big\{\delta \in \R^{d_1}: \|\delta - \tilde{\delta}(y) \| \le \eta \big\}$
		implies
		that, 
		given the event 
		$\big\{
		\sup_{y \in \mathcal{Y}}\| \hat{\delta}(y) - \tilde{\delta}(y) \|
		\ge \eta
		\big\}$,
		we have,
		for any $y \in \mathcal{Y}$,
		\begin{eqnarray}
			\label{eq:Q1}
			\sup_{\delta \in D_{n, \eta}(y)}
			Q_{n, y}
			(\delta)
			-
			Q_{n, y}
			\big (\tilde{\delta}(y)\big) \ge 0,
		\end{eqnarray}
		where
		the last inequality is due to that 
		$Q_{n, y}
		\big (\hat{\delta}(y)\big)
		-
		Q_{n, y}
		\big ( \tilde{\delta}(y)\big)
		\ge  0
		$,
		by definition of
		$\hat{\delta}(y)$.
		It follows from
		(\ref{eq:lbd1})-(\ref{eq:Q1}) that,
		given the event 
		$\big\{
		\sup_{y \in \mathcal{Y}}\| \hat{\delta}(y) - \tilde{\delta}(y) \|
		\ge \eta
		\big\}$,
		\begin{eqnarray*}
			\sup_{\delta \in D_{n, \eta}(y)}
			\big|
			Q_{n, y}
			(\delta)
			-
			\widetilde{Q}_{n, y}
			(\delta)
			\big|
			\ge
			\frac{c}{4n} \eta^{2}. 
		\end{eqnarray*}
		Because
		$\Psi_{n}
		\big (\theta_{0}(y), y\big)$  
		is Donsker by Lemma \ref{lemma:donsker},
		we can show that,
		for any $\xi>0$,
		there exists a constant $C$
		such that 
		$\Pr\big(\sup_{y \in \mathcal{Y}}\| \tilde{\delta}(y) \| \ge C\big)
		\le \xi
		$
		for sufficiently large $n$. 
		Thus, the above display implies that 
		\begin{eqnarray*}
			\Pr
			\bigg (
			\sup_{y \in \mathcal{Y}}\| \hat{\delta}(y) - \tilde{\delta}(y) \|
			\ge \eta 
			\bigg )
			\le
			\Pr 
			\bigg (
			\sup_{y \in \mathcal{Y}}
			\sup_{\delta: \|\delta \| \le \eta + C}
			\big |
			Q_{n, y}
			(\delta)
			-
			\widetilde{Q}_{n, y}
			(\delta)
			\big |
			>
			\frac{c }{4n} \eta^2
			\bigg )
			+
			\xi,
		\end{eqnarray*}
		for sufficiently large $n$.
		It follows from (\ref{eq:Q2}) that
		the first term on the right side of the above equation
		converges to 0 as $n \to \infty$.
		Thus, we obtain the desired conclusion.
	\end{proof}

	\vspace{0.5cm}
	\begin{proof}[\textbf{Proof of Proposition \ref{pro:est}}]
		
		Lemma \ref{lemma:argmin} implies that, 
		uniformly in $(y, z) \in \mathcal{Y} \times \mathcal{Z}$,
		\begin{eqnarray*}
			\sqrt{n}
			\big (
			\hat{\theta}(y,z)
			-
			\theta_{0}(y,z)
			\big )
			=
			-
			H_{0}(y, z)^{-1}
			G_{n}(y,z) 
			+ o_p(1),
		\end{eqnarray*}
		where 
		$
		G_{n}(y, z)  
		:=
		\Psi_{n}
		\big (\theta_{0}(y,z), y, z \big)$.
		By the implicit function theorem, we can show that
		$\alpha_{0}(y)$
		is differentiable uniformly over $y \in \mathcal{Y}$.
		Thus, the empirical process
		$G_{n}(y, z)$ is 
		stochastically equicontinuous over $\mathcal{Y}$
		for any $z \in \mathcal{Z}$.
		Given iid observations under Assumption A1,
		the finite dimensional convergence follows from 
		a multivariate central limit theorem.
		This with
		the stochastic equicontinuity $y \mapsto G_{n}(y,z)$
		and the finite set $\mathcal{Z}$
		imply that 
		$
		G_{n}(\cdot,\cdot)
		\rightsquigarrow
		\mathbb{G}(\cdot,\cdot)
		$
		in
		$\ell^{\infty}(\mathcal{Y})^{d_1}
		{\times}
		\ell^{\infty}(\mathcal{Z})^{d_2}$,
		where $\mathbb{G}(\cdot,\cdot)$ is
		a zero-mean Gaussian process with covariance function
		defined in Proposition \ref{pro:est}.
	\end{proof}
	\vspace{0.5cm}

	\begin{proof}[\textbf{Proof of Theorem \ref{theorem:CLT}}]
		\textbf{(a)}
		Consider the map
		$\phi:
		\mathbb{D}_{\phi}
		\subset
		\mathbb{D}
		\mapsto 
		\mathbb{S}_{\phi}
		$,
		where 
		$ (a, b) \mapsto \phi(a, b)$,
		given by 
		$
		\phi(a, b)(x,y,z)
		=
		\big[\Lambda
		\big (P_1(x,z)'a(y) \big),
		\Lambda
		\big (P_2(x)'b(z) \big) \big]'
		$.
		Under Assumption A2, 
		the map $\phi(\cdot)$
		is shown to be Hadamard differentiable
		at
		$
		\theta(\cdot)
		=
		\big(
		\alpha(\cdot)'  ,
		\beta(\cdot)' \big)'
		$
		tangentially to
		$\mathbb{D}$
		with the derivative map
		$(a, b)
		\mapsto 
		\phi_{\theta_{0}(\cdot)}'(a, b)$,
		given by 
		\begin{eqnarray*}
			\phi_{\theta_{0}(\cdot)}'(a, b)(x,y,z)
			=
			\big [
			\lambda
			\big (P_1(x,z)' \alpha_{0}(y)\big)
			\big( P_1(x,z)' a(y)\big), 
			\lambda
			\big (
			P_2(x)'\beta_{0}(z)
			\big)
			P_2(x)'b(z)
			\big ] '.
		\end{eqnarray*}
		Then,
		we can write 
		$
		\big (
		\widehat{F}_{Y|X, Z},
		\widehat{F}_{Z|X}
		\big )'
		=
		\phi\big(\hat{\theta}(\cdot) \big)
		$
		and
		$
		\big (
		F_{Y|X, Z},
		F_{Z|X}
		\big )'
		=
		\phi\big(\theta_{0}(\cdot) \big)
		$.
		Applying the functional delta method
		with
		the result in Proposition \ref{pro:est},  
		we can show that
		\begin{eqnarray*}
			\sqrt{n}
			\left (
			\begin{array}{c}
				\widehat{F}_{Y|X, Z} - F_{Y|X, Z} \\
				\widehat{F}_{Z|X} - F_{Z|X}
			\end{array}
			\right )
			\rightsquigarrow
			\phi_{\theta_{0}(\cdot)}'
			(
			\mathbb{B}
			)
			\ \ \mathrm{in} \ \
			\ell^{\infty}(\mathcal{X}{\times}\mathcal{Y}{\times}\mathcal{Z})
			{\times}
			\ell^{\infty}(\mathcal{X}{\times}\mathcal{Z}).
		\end{eqnarray*}
		
		\noindent 
		\textbf{(b)}
		The chain rule for Hadamard differentiable maps
		\citep[Lemma 3.9.3,][]{van1996weak} shows that
		$\nu \circ \phi: \mathbb{D}_{\phi} \to
		\ell^{\infty}(\mathcal{X}{\times}\mathcal{Y}{\times}\mathcal{Z})$
		is Hadamard differntiable
		at $\theta$ tangentially to $\mathbb{D}$
		with
		derivative
		$\nu'_{\phi(\theta)} \circ \phi_{\theta}'$.
		An application of the functional delta method yields
		the desired conclusion.
	\end{proof}

	\vspace{0.5cm}
	\begin{proof}
		[\textbf{Proof of Theorem \ref{theorem:bootstrap}}]
		
		\textbf{(a)}
		Define 
		$\Psi_{n}^{\ast}(\theta, y, z):=
		n^{-1/2}
		\sum_{i=1}^{n}
		\big[
		\nabla \ell_{i, y}^{\ast}
		\big (
		\alpha
		\big)',
		\nabla \ell_{i, z}^{\ast}
		\big (
		\beta
		\big)'\big]'$
		and
		let
		$
		G_{n}^{\ast}(y, z)
		:=
		\Psi_{n}^{\ast}
		\big (\theta_{0}(y,z), y, z \big)$.
		Applying a similar argument used in Lemma \ref{lemma:argmin},
		we can show that,
		uniformly in $(y, z) \in \mathcal{Y} \times \mathcal{Z}$,
		\begin{eqnarray*}
			\sqrt{n}
			\big (
			\hat{\theta}^{\ast}(y, z)
			-
			\theta_{0}(y, z)
			\big )
			=
			-
			H_{0, y}^{-1}
			G_{n}^{\ast}(y,z)
			+ o_p(1).
		\end{eqnarray*}
		This together with 
		Proposition \ref{pro:est}
		yields,
		uniformly in $(y, z) \in \mathcal{Y} \times \mathcal{Z}$,
		\begin{eqnarray*}
			\sqrt{n}
			\big (
			\hat{\theta}^{\ast}(y, z)
			-
			\hat{\theta}(y, z)
			\big )
			=
			-
			H_{0, y}^{-1}
			\big[
			G_{n}^{\ast}(y,z)
			-
			G_{n}(y,z)
			\big]
			+ o_p(1).
		\end{eqnarray*}
		Because 
		the class of gradient functions is Donsker
		from Lemma \ref{lemma:donsker},
		Theorem 3.6.13 of \cite{van1996weak} implies that   
		$
		G_{n}^{\ast}(\cdot,\cdot)
		-
		G_{n}(\cdot,\cdot)
		\rightsquigarrow^{p}
		\mathbb{G}(\cdot,\cdot)
		$
		in
		$\ell^{\infty}(\mathcal{Y} {\times} \mathcal{Z})$.
		It follows that 
		$
		\sqrt{n}
		\big (
		\hat{\theta}^{\ast}(\cdot,\cdot)
		-
		\hat{\theta}(\cdot,\cdot)
		\big )
		\rightsquigarrow^{p}
		\mathbb{B}(\cdot,\cdot)
		$
		in 
		$     \ell^{\infty}(\mathcal{Y})^{d_1}
		{\times}
		\ell^{\infty}(\mathcal{Z})^{d_2}.
		$
		Applying the functional delta method,
		we can show 
		\begin{eqnarray*}
			\sqrt{n}
			\left (
			\begin{array}{c}
				\widehat{F}_{Y|X, Z}^{\ast} - \widehat{F}_{Y|X, Z} \\
				\widehat{F}_{Z|X}^{\ast} - \widehat{F}_{Z|X}
			\end{array}
			\right )
			\rightsquigarrow^{p}
			\phi_{\theta_{0}(\cdot)}'
			(
			\mathbb{B}
			)
			\ \ \mathrm{in} \ \
			\ell^{\infty}(\mathcal{X}{\times}\mathcal{Y}{\times}\mathcal{Z})
			{\times}
			\ell^{\infty}(\mathcal{X}{\times}\mathcal{Z}).
		\end{eqnarray*}
		
		\noindent 
		\textbf{(b)}
		Also,
		for a Hadamard differentiable map
		$\nu(\cdot)$,
		the functional delta method leads to 
		\begin{eqnarray*}
			\sqrt{n}
			\big \{
			\nu
			\big(
			\widehat{F}_{Y|X, Z}^{\ast},
			\widehat{F}_{Z|X}^{\ast}
			\big) 
			-
			\nu(
			\widehat{F}_{Y|X, Z},
			\widehat{F}_{Z|X}) 
			\big \}
			\rightsquigarrow^{p}
			\nu_{F_{Y|X,Z}, F_{Z|X}}'
			\circ
			\phi_{\theta_{0}(\cdot)}'
			\big (
			\mathbb{B}
			\big ),
		\end{eqnarray*}
		in
		$  \ell^{\infty}(\mathcal{X}{\times}\mathcal{Y}{\times}\mathcal{Z}) $.
	\end{proof}
	
	\newpage
	
	\section{Additional Simulation Results}\label{sec: appendix-B}
	
	We provide additional simulation results that further comfirm the conclusion we obtained in Section \ref{sec:simulation}. First, the 95th conditional quantile, mean, and standard deviation of $Y$ under each DGP are given in Tables \ref{tab:PGB2-Y}, \ref{tab: Copula-Y} and \ref{tab:Normal-Y}. Besides, another DGP based on the hierarchical model with Negative Binomial and Log Normal distributions is considered, the results of $C$ and $Y$ are presented in Tables \ref{tab: NB-C} and \ref{tab: NB-Y}, respectively. 
	
	All the comparisons in Section \ref{sec:simulation} are conducted by looking at three different covariate values. Here, we provide additional comparion results by looking at $1000$ randomly generated covariates from the uniform distributions, which is considered as a cohort. For all the four DGPs, the 95\% VaR and ES of $C$ for this cohort are explored in both cases. Specifically, in Tables \ref{tab: VaR-C} and \ref{tab: ES-C}, the estimated values and the 95\% confidence interval obtained based on $1000$ monte carlo simulations are given for each risk measure. For both measures, the DR approach outperforms the parametric models in most cases, like under the Truncated Normal and NB-LN hierarchical DGPs, with slightly wider but more confident confidence intervals and more accurate point estimates. In some cases, such as the Copula DGP, the correctly specificed parametric model gives the best results, but the DR approach performs comparatively.

	\begin{table}[H] 
		\centering
		\caption{Monte Carlo Results under Poisson-GB2 Hierarchical DGP\label{tab:PGB2-Y}}
		\linespread{1.0}
		\small
		\begin{tabular}{llrrrrrrrrr}
			\hline
			&                        & \multicolumn{1}{l}{}     & \multicolumn{1}{l}{} & \multicolumn{3}{c}{Bias}                                                   & \multicolumn{1}{l}{} & \multicolumn{3}{c}{MSE}                                                    \\ \cline{5-7} \cline{9-11} 
			Quantities                         &                        & $x_1$                    & \multicolumn{1}{l}{} & \multicolumn{1}{c}{DR} & \multicolumn{1}{c}{H} & \multicolumn{1}{c}{Copula} & \multicolumn{1}{l}{} & \multicolumn{1}{c}{DR} & \multicolumn{1}{c}{H} & \multicolumn{1}{c}{Copula} \\ \hline
			\multirow{6}{*}{$Q_{0.95}(Y|x)$}   & {Case1} & 0.25                     &                      & -0.06                  & -0.01                   & 0.02                    &                      & 0.11                   & 0.04                    & 0.05                    \\
			&                        & 0.5                      &                      & 0.01                   & 0.05                    & 0.19                    &                      & 0.21                   & 0.06                    & 0.13                    \\
			&                        & 0.75                     &                      & -0.02                  & 0.01                    & 0.27                    &                      & 0.44                   & 0.10                    & 0.22                    \\
			& {Case2} & 0.25                     &                      & -0.05                  & -0.02                   & 0.16                    &                      & 0.11                   & 0.03                    & 0.08                    \\
			&                        & 0.5                      &                      & 0.00                   & 0.04                    & 0.18                    &                      & 0.20                   & 0.06                    & 0.12                    \\
			&                        & 0.75                     &                      & -0.01                  & 0.06                    & 0.08                    &                      & 0.34                   & 0.10                    & 0.14                    \\ \hline
			\multirow{6}{*}{$\mathbb{E}(Y|x)$} & {Case1} & 0.25                     &                      & 0.04                   & 0.02                    & 0.05                    &                      & 0.02                   & 0.01                    & 0.01                    \\
			&                        & 0.5                      &                      & 0.04                   & 0.02                    & 0.05                    &                      & 0.04                   & 0.01                    & 0.02                    \\
			&                        & 0.75                     &                      & 0.04                   & 0.01                    & 0.04                    &                      & 0.05                   & 0.02                    & 0.02                    \\
			& {Case2} & \multicolumn{1}{l}{0.25} & \multicolumn{1}{l}{} & -0.01                  & -0.03                   & 0.01                    &                      & 0.01                   & 0.01                    & 0.01                    \\
			&                        & 0.5                      &                      & -0.01                  & -0.02                   & 0.01                    &                      & 0.03                   & 0.01                    & 0.01                    \\
			&                        & 0.75                     &                      & 0.06                   & 0.04                    & 0.07                    &                      & 0.05                   & 0.02                    & 0.02                    \\ \hline
			\multirow{6}{*}{$Std(Y|x)$}        & {Case1} & 0.25                     &                      & -0.002                 & -0.001                  & 0.01                    &                      & 0.01                   & 0.01                    & 0.01                    \\
			&                        & 0.5                      &                      & -0.04                  & -0.04                   & 0.00                    &                      & 0.03                   & 0.01                    & 0.01                    \\
			&                        & 0.75                     &                      & -0.03                  & -0.01                   & 0.06                    &                      & 0.05                   & 0.01                    & 0.02                    \\
			& {Case2} & 0.25                     &                      & -0.02                  & -0.02                   & 0.03                    &                      & 0.01                   & 0.01                    & 0.01                    \\
			&                        & 0.5                      &                      & -0.04                  & -0.03                   & 0.01                    &                      & 0.02                   & 0.01                    & 0.01                    \\
			&                        & 0.75                     &                      & -0.04                  & -0.01                   & -0.01                   &                      & 0.03                   & 0.01                    & 0.02                    \\ \hline
		\end{tabular}
		\begin{minipage}{0.82\linewidth} \linespread{1.0}\footnotesize
			\textit{Notes}:
			The number of
			Monte Carlo iterations
			is set to 1,000.
			We choose covariates $x=(1, x_1,0.5,0.5)$ with $x_1\in\{0.25, 0.50, 0.75\}$.
			For each quantity, we report the bias and MSE in both cases. For simplicity, we represent the hierarchical and Gaussian copula models as `H' and `Copula', respectively.
		\end{minipage}
	\end{table}

	\begin{table}[H] 
		\centering
		\caption{Monte Carlo Results under Poisson-Gamma Gaussian Copula DGP\label{tab: Copula-Y}}
		\linespread{1.0}
		\small
		\begin{tabular}{llrrrrrrrrr}
			\hline
			&                        & \multicolumn{1}{l}{}     & \multicolumn{1}{l}{} & \multicolumn{3}{c}{Bias}                                                   & \multicolumn{1}{l}{} & \multicolumn{3}{c}{MSE}                                                    \\ \cline{5-7} \cline{9-11} 
			Quantities                         &                        & $x_1$                    & \multicolumn{1}{l}{} & \multicolumn{1}{c}{DR} & \multicolumn{1}{c}{H} & \multicolumn{1}{c}{Copula} & \multicolumn{1}{l}{} & \multicolumn{1}{c}{DR} & \multicolumn{1}{c}{H} & \multicolumn{1}{c}{Copula} \\ \hline
			\multirow{6}{*}{$Q_{0.95}(Y|x)$}   & {Case1} & 0.25                     &                      & 0.09                   & -0.63                   & -0.17                   &                      & 0.08                   & 0.44                    & 0.07                    \\
			&                        & 0.5                      &                      & 0.03                   & -0.87                   & -0.13                   &                      & 0.11                   & 0.84                    & 0.10                    \\
			&                        & 0.75                     &                      & -0.04                  & -1.10                   & -0.06                   &                      & 0.23                   & 1.37                    & 0.15                    \\
			& {Case2} & 0.25                     &                      & 0.08                   & -0.41                   & -0.03                   &                      & 0.12                   & 0.23                    & 0.07                    \\
			&                        & 0.5                      &                      & 0.05                   & -0.53                   & 0.00                    &                      & 0.19                   & 0.40                    & 0.11                    \\
			&                        & 0.75                     &                      & 0.04                   & -0.68                   & 0.06                    &                      & 0.37                   & 0.66                    & 0.19                    \\ \hline
			\multirow{6}{*}{$\mathbb{E}(Y|x)$} & {Case1} & 0.25                     &                      & -0.19                  & 0.05                    & 0.16                    &                      & 0.07                   & 0.04                    & 0.05                    \\
			&                        & 0.5                      &                      & -0.28                  & 0.01                    & 0.27                    &                      & 0.15                   & 0.07                    & 0.12                    \\
			&                        & 0.75                     &                      & -0.38                  & -0.10                   & 0.39                    &                      & 0.30                   & 0.14                    & 0.25                    \\
			& {Case2} & \multicolumn{1}{l}{0.25} & \multicolumn{1}{l}{} & -0.09                  & 0.00                    & 0.21                    &                      & 0.06                   & 0.04                    & 0.07                    \\
			&                        & 0.5                      &                      & -0.09                  & -0.02                   & 0.28                    &                      & 0.09                   & 0.07                    & 0.12                    \\
			&                        & 0.75                     &                      & -0.08                  & -0.06                   & 0.37                    &                      & 0.15                   & 0.13                    & 0.20                    \\ \hline
			\multirow{6}{*}{$Std(Y|x)$}        & {Case1} & 0.25                     &                      & 0.14                   & -0.26                   & -0.12                   &                      & 0.02                   & 0.07                    & 0.02                    \\
			&                        & 0.5                      &                      & 0.17                   & -0.37                   & -0.14                   &                      & 0.04                   & 0.14                    & 0.02                    \\
			&                        & 0.75                     &                      & 0.20                   & -0.46                   & -0.17                   &                      & 0.05                   & 0.22                    & 0.03                    \\
			& {Case2} & 0.25                     &                      & 0.08                   & -0.25                   & -0.09                   &                      & 0.02                   & 0.06                    & 0.01                    \\
			&                        & 0.5                      &                      & 0.06                   & -0.32                   & -0.11                   &                      & 0.02                   & 0.11                    & 0.02                    \\
			&                        & 0.75                     &                      & 0.05                   & -0.41                   & -0.12                   &                      & 0.03                   & 0.18                    & 0.02                    \\ \hline		
		\end{tabular}
		\begin{minipage}{.8\linewidth} \linespread{1.0}\footnotesize
			\textit{Notes}: Refer to Table \ref{tab:PGB2-Y}.
		\end{minipage}
	\end{table}

	\begin{table}[H] 
		\centering
		\caption{Monte Carlo Results under Truncated Bivariate Normal DGP\label{tab:Normal-Y}}
		\linespread{1.0}
		\small
		\begin{tabular}{llrrrrrrrrr}
			\hline
			&                        & \multicolumn{1}{l}{}     & \multicolumn{1}{l}{} & \multicolumn{3}{c}{Bias}                                                   & \multicolumn{1}{l}{} & \multicolumn{3}{c}{MSE}                                                    \\ \cline{5-7} \cline{9-11} 
			Quantities                         &                        & $x_1$                    & \multicolumn{1}{l}{} & \multicolumn{1}{c}{DR} & \multicolumn{1}{c}{H} & \multicolumn{1}{c}{Copula} & \multicolumn{1}{l}{} & \multicolumn{1}{c}{DR} & \multicolumn{1}{c}{H} & \multicolumn{1}{c}{Copula} \\ \hline
			\multirow{6}{*}{$Q_{0.95}(Y|x)$}   & {Case1} & 0.25                     &                      & -0.29                  & 0.30                    & -2.78                   &                      & 0.87                   & 1.14                    & 8.05                    \\
			&                        & 0.5                      &                      & -0.37                  & 0.58                    & -2.85                   &                      & 1.03                   & 1.68                    & 8.49                    \\
			&                        & 0.75                     &                      & -0.42                  & 0.81                    & -2.96                   &                      & 1.05                   & 2.07                    & 9.08                    \\
			& {Case2} & 0.25                     &                      & -0.06                  & -0.77                   & 0.43                    &                      & 0.17                   & 0.74                    & 0.35                    \\
			&                        & 0.5                      &                      & 0.06                   & -0.67                   & 0.52                    &                      & 0.20                   & 0.60                    & 0.46                    \\
			&                        & 0.75                     &                      & 0.19                   & -0.49                   & 0.64                    &                      & 0.22                   & 0.38                    & 0.57                    \\ \hline
			\multirow{6}{*}{$\mathbb{E}(Y|x)$} &{Case1} & 0.25                     &                      & -0.12                  & 0.30                    & 0.11                    &                      & 0.09                   & 0.20                    & 0.09                    \\
			&                        & 0.5                      &                      & -0.17                  & 0.34                    & 0.09                    &                      & 0.12                   & 0.25                    & 0.11                    \\
			&                        & 0.75                     &                      & -0.18                  & 0.35                    & 0.05                    &                      & 0.13                   & 0.26                    & 0.09                    \\
			& {Case2} & \multicolumn{1}{l}{0.25} & \multicolumn{1}{l}{} & 0.02                   & -0.02                   & 0.38                    &                      & 0.02                   & 0.02                    & 0.17                    \\
			&                        & 0.5                      &                      & 0.08                   & 0.03                    & 0.44                    &                      & 0.03                   & 0.02                    & 0.22                    \\
			&                        & 0.75                     &                      & 0.08                   & 0.03                    & 0.44                    &                      & 0.03                   & 0.02                    & 0.22                    \\ \hline
			\multirow{6}{*}{$Std(Y|x)$}        &{Case1} & 0.25                     &                      & -0.16                  & 1.04                    & -0.92                   &                      & 0.08                   & 1.39                    & 0.88                    \\
			&                        & 0.5                      &                      & -0.17                  & 1.21                    & -0.95                   &                      & 0.09                   & 1.84                    & 0.95                    \\
			&                        & 0.75                     &                      & -0.19                  & 1.38                    & -1.00                   &                      & 0.09                   & 2.35                    & 1.02                    \\
			& {Case2} & 0.25                     &                      & 0.00                   & -0.05                   & 0.33                    &                      & 0.02                   & 0.03                    & 0.14                    \\
			&                        & 0.5                      &                      & 0.04                   & -0.02                   & 0.35                    &                      & 0.02                   & 0.03                    & 0.15                    \\
			&                        & 0.75                     &                      & 0.09                   & 0.05                    & 0.39                    &                      & 0.03                   & 0.03                    & 0.18                    \\ \hline
		\end{tabular}
		\begin{minipage}{.75\linewidth} \linespread{1.0}\footnotesize
			\textit{Notes}: Refer to Table \ref{tab:PGB2-Y}.
		\end{minipage}
	\end{table}

	\begin{table}[H]
		\centering
		\caption{ Monte Carlo Results under NB-LN Hierarchical DGP\label{tab: NB-C}}
		\linespread{1.0}	
		\small	
		\begin{tabular}{llrrrrrrrrr}
			\hline\hline
			&       & \multicolumn{1}{l}{}      &  & \multicolumn{3}{c}{Bias}                                                    & \multicolumn{1}{c}{} & \multicolumn{3}{c}{MSE}                                                     \\ \cline{5-7} \cline{9-11} 
			Quantities                         & Case  & \multicolumn{1}{r}{$x_1$} &  & \multicolumn{1}{c}{DR} & \multicolumn{1}{c}{H} & \multicolumn{1}{c}{Copula} & \multicolumn{1}{c}{} & \multicolumn{1}{c}{DR} & \multicolumn{1}{c}{H} & \multicolumn{1}{c}{Copula} \\ \hline
			\multirow{6}{*}{$\mathbb{E}(C|x)$} & Case1 & 0.25                      &  & 0.07                   & 0.001                 & -0.45                      &                      & 0.02                   & 0.009                 & 0.21                       \\
			&       & 0.50                      &  & 0.01                   & -0.050                & -0.62                      &                      & 0.02                   & 0.015                 & 0.39                       \\
			&       & 0.75                      &  & 0.03                   & -0.024                & -0.76                      &                      & 0.03                   & 0.017                 & 0.59                       \\
			& Case2 & 0.25                      &  & -0.06                  & -0.95                 & -1.03                      &                      & 0.01                   & 0.91                  & 1.06                       \\
			&       & 0.50                      &  & 0.02                   & -0.87                 & -0.94                      &                      & 0.01                   & 0.77                  & 0.89                       \\
			&       & 0.75                      &  & 0.02                   & -0.86                 & -0.93                      &                      & 0.01                   & 0.74                  & 0.86                       \\ \hline
			\multirow{6}{*}{$Std(C|x)$}        & Case1 & 0.25                      &  & 0.11                   & 0.00                  & 0.48                       &                      & 0.02                   & 0.002                 & 0.24                       \\
			&       & 0.50                      &  & 0.10                   & -0.03                 & 0.26                       &                      & 0.02                   & 0.005                 & 0.07                       \\
			&       & 0.75                      &  & 0.14                   & 0.05                  & 0.05                       &                      & 0.04                   & 0.010                 & 0.01                       \\
			& Case2 & 0.25                      &  & -0.05                  & -0.82                 & -0.92                      &                      & 0.01                   & 0.67                  & 0.86                       \\
			&       & 0.50                      &  & 0.04                   & -0.87                 & -0.99                      &                      & 0.02                   & 0.76                  & 0.98                       \\
			&       & 0.75                      &  & 0.00                   & -1.01                 & -1.17                      &                      & 0.03                   & 1.04                  & 1.37                       \\ \hline
			\multirow{6}{*}{$ES_{0.95}(C|x)$}  & Case1 & 0.25                      &  & 0.81                   & 0.09                  & 2.04                       &                      & 0.79                   & 0.05                  & 4.23                       \\
			&       & 0.50                      &  & 0.65                   & -0.13                 & 0.87                       &                      & 0.64                   & 0.10                  & 0.89                       \\
			&       & 0.75                      &  & 0.82                   & 0.17                  & 0.00                       &                      & 1.10                   & 0.16                  & 0.19                       \\
			& Case2 & 0.25                      &  & -0.24                  & -2.14                 & -2.45                      &                      & 0.21                   & 4.64                  & 6.08                       \\
			&       & 0.50                      &  & 0.09                   & -2.20                 & -2.69                      &                      & 0.28                   & 4.96                  & 7.36                       \\
			&       & 0.75                      &  & 0.01                   & -2.70                 & -3.41                      &                      & 0.48                   & 7.45                  & 11.81                      \\ \hline
			\multirow{6}{*}{$Q_{0.95}(C|x)$}   & Case1 & 0.25                      &  & 0.04                   & -0.03                 & 1.59                       &                      & 0.04                   & 0.02                  & 2.59                       \\
			&       & 0.50                      &  & -0.06                  & -0.08                 & 0.98                       &                      & 0.09                   & 0.04                  & 1.05                       \\
			&       & 0.75                      &  & 0.09                   & 0.08                  & 0.34                       &                      & 0.20                   & 0.07                  & 0.22                       \\
			& Case2 & 0.25                      &  & -0.18                  & -1.88                 & -2.58                      &                      & 0.11                   & 3.56                  & 6.69                       \\
			&       & 0.50                      &  & -0.13                  & -2.26                 & -3.00                      &                      & 0.15                   & 5.16                  & 9.05                       \\
			&       & 0.75                      &  & 0.08                   & -2.66                 & -3.45                      &                      & 0.26                   & 7.18                  & 12.00                      \\ \hline
		\end{tabular}
		\begin{minipage}{.8\linewidth} \footnotesize
			\textit{Notes}: Refer to Table \ref{tab:PGB2-Y}.
		\end{minipage}
	\end{table}
	
	\begin{table}[H] 
		\centering
		\caption{Monte Carlo Results under NB-LN Hierarchical DGP\label{tab: NB-Y}}
		\linespread{1.0}
		\small
		\begin{tabular}{llrrrrrrrrr}
			\hline\hline
			&       & \multicolumn{1}{l}{}      &  & \multicolumn{3}{c}{Bias}                                                    & \multicolumn{1}{c}{} & \multicolumn{3}{c}{MSE}                                                     \\ \cline{5-7} \cline{9-11} 
			Quantities                         & Case  & \multicolumn{1}{r}{$x_1$} &  & \multicolumn{1}{c}{DR} & \multicolumn{1}{c}{H} & \multicolumn{1}{c}{Copula} & \multicolumn{1}{c}{} & \multicolumn{1}{c}{DR} & \multicolumn{1}{c}{H} & \multicolumn{1}{c}{Copula} \\ \hline
			\multirow{6}{*}{$\mathbb{E}(Y|x)$} & Case1 & 0.25                      &  & -0.002                 & -0.0002               & 0.17                       &                      & 0.006                  & 0.006                 & 0.03                       \\
			&       & 0.50                      &  & -0.002                 & -0.0009               & 0.29                       &                      & 0.010                  & 0.010                 & 0.09                       \\
			&       & 0.75                      &  & 0.001                  & -0.0010               & 0.46                       &                      & 0.016                  & 0.016                 & 0.22                       \\
			& Case2 & 0.25                      &  & 0.02                   & 0.0002                & 1.42                       &                      & 0.04                   & 0.04                  & 2.02                       \\
			&       & 0.50                      &  & 0.05                   & 0.0001                & 1.97                       &                      & 0.07                   & 0.07                  & 3.91                       \\
			&       & 0.75                      &  & 0.09                   & 0.0007                & 2.66                       &                      & 0.12                   & 0.11                  & 7.11                       \\ \hline
			\multirow{6}{*}{$Std(Y|x)$}        & Case1 & 0.25                      &  & -0.003                 & 0.001                 & -0.75                      &                      & 0.01                   & 0.01                  & 0.56                       \\
			&       & 0.50                      &  & -0.010                 & -0.002                & -0.90                      &                      & 0.01                   & 0.01                  & 0.81                       \\
			&       & 0.75                      &  & -0.004                 & -0.001                & -1.05                      &                      & 0.02                   & 0.01                  & 1.11                       \\
			& Case2 & 0.25                      &  & -0.015                 & 0.002                 & -0.03                      &                      & 0.01                   & 0.01                  & 0.01                       \\
			&       & 0.50                      &  & -0.023                 & 0.002                 & 0.03                       &                      & 0.02                   & 0.01                  & 0.01                       \\
			&       & 0.75                      &  & -0.031                 & 0.002                 & 0.10                       &                      & 0.03                   & 0.02                  & 0.03                       \\ \hline
			\multirow{6}{*}{$Q_{0.95}(Y|x)$}   & Case1 & 0.25                      &  & 0.003                  & -0.0002               & -1.373                     &                      & 0.06                   & 0.06                  & 1.89                       \\
			&       & 0.50                      &  & 0.006                  & -0.0075               & -1.578                     &                      & 0.11                   & 0.10                  & 2.51                       \\
			&       & 0.75                      &  & 0.029                  & -0.0034               & -1.793                     &                      & 0.19                   & 0.16                  & 3.24                       \\
			& Case2 & 0.25                      &  & -0.006                 & 0.004                 & 1.17                       &                      & 0.14                   & 0.11                  & 1.43                       \\
			&       & 0.50                      &  & 0.034                  & 0.005                 & 1.78                       &                      & 0.26                   & 0.18                  & 3.29                       \\
			&       & 0.75                      &  & 0.084                  & 0.013                 & 2.57                       &                      & 0.47                   & 0.31                  & 6.86                       \\ \hline
		\end{tabular}
		\begin{minipage}{.85\linewidth} \linespread{1.0}\footnotesize
			\textit{Notes}: Refer to Table \ref{tab:PGB2-Y}.
		\end{minipage}
	\end{table}

	\begin{table}[H] 
		\centering
		\caption{$VaR_{0.95}(C|x)$ for a Cohort\label{tab: VaR-C}}
		\linespread{1.0}
		\small
		\begin{tabular}{llcccc}
			\hline\hline
			DGPs                         & Case  & Empirical            & DR                                 & H                   & Copula                             \\ \hline
			H (P-GB2)        & Case1 & 13.33                & 14.15                              & 14.01                              & 12.19                              \\
			&       &                      & (13.00, 15.35)                     & (12.99, 15.03)                     & (11.22, 13.13)                     \\
			& Case2 & 16.18                & 17.00                              & 16.83                              & 16.23                              \\
			&       & \multicolumn{1}{l}{} & \multicolumn{1}{l}{(15.80, 18.40)} & \multicolumn{1}{l}{(15.70, 18.04)} & \multicolumn{1}{l}{(14.75, 17.86)} \\ \hline
			Copula (P-G)        & Case1 & 72.16                & 72.30                              & 70.57                              & 72.86                              \\
			&       &                      & (66.23, 79.63)                     & (65.38, 76.54)                     & (66.20, 80.54)                     \\
			& Case2 & 13.46                & 15.20                              & 15.11                              & 13.88                              \\
			&       & \multicolumn{1}{l}{} & (13.5, 17.23)                      & (13.38, 16.96)                     & (12.25, 15.65)                     \\ \hline
			Normal & Case1 & 35.74                & 37.40                              & 46.32                              & 30.97                              \\
			&       & \multicolumn{1}{l}{} & \multicolumn{1}{l}{(33.60, 41.10)} & \multicolumn{1}{l}{(39.38, 58.01)} & \multicolumn{1}{l}{(28.02, 33.86)} \\
			& Case2 & 15.48                & 17.05                              & 18.8                               & 18.29                              \\
			&       & \multicolumn{1}{l}{} & \multicolumn{1}{l}{(14.90, 21.83)} & \multicolumn{1}{l}{(15.42, 22.73)} & \multicolumn{1}{l}{(16.32, 20.88)} \\ \hline
			H (NB-LN)          & Case1 & 16.26                & 16.15                              & 15.21                              & 14.99                              \\
			&       &                      & (14.80, 17.55)                     & (14.32, 16.22)                     & (14.09, 16.02)                     \\
			& Case2 & 9.15                 & 8.90                               & 8.61                               & 7.26                               \\
			&       &                      & (8.00, 10.00)                      & (7.74, 9.49)                       & (6.54, 8.10)                       \\ \hline
		\end{tabular}
		\begin{minipage}{0.82\linewidth} \linespread{1.0}\footnotesize
			\textit{Notes}:
			The $VaR_{0.95}(C|x)$ is explored across a cohort that constructed by randomly generating 1,000 covariates $x$ from the uniform distribution. For each scenario, the empirical values are the $VaR_{0.95}(C|x)$ based on the observations generated from the true DGP. For the three estimation models, the mean estimates (individual numbers) and the corresponding $95\%$ confidence intervals (bracketed numbers) based on 1,000 Monte Carlo iterations are presented.
		\end{minipage}   
	\end{table}

	\begin{table}[H] 
		\centering
		\caption{$ES_{0.95}(C|x)$ for a Cohort\label{tab: ES-C}}
		\linespread{1.0}
		\small
		\begin{tabular}{llcccc}
			\hline\hline
			DGPs         & Case  & Empirical            & DR                                 & Hierarchical                        & Copula                             \\ \hline
			H (P-GB2)    & Case1 & 17.35                & 18.93                              & 18.56                               & 16.17                              \\
			&       &                      & (17.28, 21.27)                     & (17.09, 20.54)                      & (14.74, 17.91)                     \\
			& Case2 & 22.36                & 22.55                              & 21.87                               & 22.92                              \\
			&       & \multicolumn{1}{l}{} & \multicolumn{1}{l}{(20.62, 25.22)} & \multicolumn{1}{l}{(20.20, 23.69)}  & \multicolumn{1}{l}{(20.58, 25.80)} \\ \hline
			Copula (P-G) & Case1 & 120.69               & 109.90                             & 100.24                              & 115.82                             \\
			&       &                      & (99.34, 122.88)                    & (92.68, 109.46)                     & (103.59, 130.40)                   \\
			& Case2 & 21.46                & 23.22                              & 22.69                               & 22.50                              \\
			&       &                      & (20.34, 27.17)                     & (20.06, 25.85)                      & (19.47, 26.53)                     \\ \hline
			Normal       & Case1 & 49.02                & 50.61                              & 95.08                               & 42.46                              \\
			&       & \multicolumn{1}{l}{} & \multicolumn{1}{l}{(45.89, 56.08)} & \multicolumn{1}{l}{(76.10, 120.92)} & \multicolumn{1}{l}{(38.35, 47.13)} \\
			& Case2 & 26.05                & 28.89                              & 33.11                               & 27.76                              \\
			&       & \multicolumn{1}{l}{} & \multicolumn{1}{l}{(24.65, 33.03)} & \multicolumn{1}{l}{(27.96, 40.70)}  & \multicolumn{1}{l}{(24.43, 31.56)} \\ \hline
			H (NB-LN)    & Case1 & 21.48                & 21.66                              & 19.31                               & 18.43                              \\
			&       &                      & (19.60, 24.56)                     & (17.85,  20.97)                     & (17.05, 19.98)                     \\
			& Case2 & 12.71                & 13.00                              & 11.98                               & 10.06                              \\
			&       &                      & (11.50, 15.10)                     & (10.64, 13.72)                      & (8.96, 11.54)                      \\ \hline
		\end{tabular}
		\begin{minipage}{.85\linewidth} \linespread{1.0}\footnotesize
			\textit{Notes}: Refer to Table \ref{tab: VaR-C}.
		\end{minipage}
	\end{table}
	
	\clearpage
	\ifodd\value{page}\else\thispagestyle{empty}\fi
	
\end{appendices}

\end{document}